\documentclass[12pt,a4paper]{article}
\usepackage{jcappub}

\usepackage{caption}
\usepackage{subfigure}
\def\l{\left}
\def\r{\right}
\def\f{\frac}
\def\ns{n_{_S}}
\def\ri{{\rm i}}

\newcommand{\impl}{\quad\Rightarrow\quad}
\def\cP{{\cal P}}
\def\cE{{\cal E}}

\usepackage{xcolor} 

\title{Tabletop potentials for inflation from $f (R)$ gravity}

\author[a,b]{Yuri Shtanov,}
\author[c]{Varun Sahni}
\author[d]{and Swagat S. Mishra}

\affiliation[a]{Bogolyubov Institute for Theoretical Physics, \\ Metrologichna St.~14-b, Kiev 03143, Ukraine}

\affiliation[b]{Astronomical Observatory, Taras Shevchenko National University of Kyiv, \\ Observatorna St.\@ 3, Kyiv 04053, Ukraine} %

\affiliation[c]{Inter-University Centre for Astronomy and Astrophysics, \\
Post Bag 4, Ganeshkhind, Pune 411~007, India}

\affiliation[d]{Centre for Astronomy and Particle Theory, School of Physics and Astronomy, University Park Campus, University of Nottingham, Nottingham NG7~2RD, UK.}

\emailAdd{shtanov@bitp.kiev.ua}
\emailAdd{varun@iucaa.in}
\emailAdd{swagat.mishra@nottingham.ac.uk}

\abstract{We show that a large class of modified gravity theories (MOG) with the Jordan-frame Lagrangian $f(R)$ translate into scalar-field (scalaron) models with hilltop potentials in the Einstein frame. (A rare exception to this rule is provided by the Starobinsky model for which the corresponding scalaron potential is plateau-like for $\phi > 0$.) We find that MOG models featuring two distinct mass scales lead to scalaron potentials that have a {\em flattened\/} hilltop, or {\em tabletop\/}. Inflationary evolution in tabletop models agrees very well with CMB observations. Tabletop potentials therefore provide a new and compelling class of MOG-based inflationary models. By contrast, MOG models with a single mass scale generally correspond to steep hilltop potentials and fail to reproduce the CMB power spectrum. Inflationary evolution in hilltop/tabletop models can proceed in two alternative directions: towards the stable point at small $R$ describing the observable universe, or towards the asymptotic region at large $R$. The MOG models which we examine have several new properties including the fact that gravity can become asymptotically vanishing, with $G_{\rm eff} \to 0$, at infinite or large finite values of the scalar curvature $R$. A universe evolving towards the asymptotically vanishing gravity region at large $R$ will either run into a `Big-Rip' singularity, or inflate eternally. }

\keywords{modified gravity, inflation}

\begin{document}
\maketitle
\flushbottom

\section{Introduction}

Metric $f (R)$ theories of gravity\footnote{In metric $f(R)$ theories, the equations of
motion are obtained by variation of the gravitational action $S_g$ with respect to the
metric $g_{\mu\nu}$, with the affine connection $\Gamma^\lambda_{\mu\nu}$ being determined 
by $g_{\mu\nu}$. In the Palatini formulation, by contrast, both $g_{\mu\nu}$ and
$\Gamma^\lambda_{\mu\nu}$ are independent variables, and the equations of motion are obtained by
varying $S_g$ with respect to both $g_{\mu\nu}$ and $\Gamma^\lambda_{\mu\nu}$. General 
relativity is distinguished by the fact that the metric and Palatini formalisms give
identical results.}
 have been extensively studied in the past as simple stable 
modified theories of gravity with one extra scalar degree of freedom, called the scalaron (see \cite{Sotiriou:2008rp, DeFelice:2010aj, Nojiri:2017ncd, tsuji_amendola} for reviews). The scalaron can be used to describe either inflation \cite{Starobinsky:1980te} or dark matter 
and dark energy \cite{Capozziello:2006uv, hu_sawicki07, star07,apple07,tsuji08,miranda09, Nojiri:2008nt, Cembranos:2008gj, Corda:2011aa, Katsuragawa:2016yir, Katsuragawa:2017wge, Yadav:2018llv, Parbin:2020bpp, KumarSharma:2022qdf, Shtanov:2021uif, Shtanov:2022wpr, Shtanov:2022xew}. The merit of $f(R)$ theories is that the scalaron's interactions with matter fields can be naturally fixed by the requirement of minimal coupling of the matter fields to gravity. 

A prototype of the inflationary $f (R)$ theory is the Starobinsky model \cite{Starobinsky:1980te, Vilenkin:1985md}, which contains only linear (Einstein) and quadratic terms in the scalar curvature, $f (R) = R + R^2 / m^2$. This is one of the most favourable inflationary models in light of the latest  observational data \cite{Planck:2018jri}. A more general analytic $f (R)$ Lagrangian, however, can be expected to contain terms of higher powers of $R$ in its expansion.  In this paper, we would like to examine systematically the types of models arising in this general case and to study their cosmological behaviour.

The problem at hand can be transparently analysed in the so-called Einstein frame.
 The Einstein frame differs from the original Jordan frame by a conformal transformation of 
the metric, in which the scalaron degree of freedom is identified as a separate scalar field $\phi$ minimally coupled to the metric governed by the usual Einstein equations. 
Different $f(R)$ theories then differ only in the form of the scalaron potential $V (\phi)$. 

We shall consider stable theories for which $f (R)$ is convex, so that $f'' (R) > 0$ and $f' (R) > 0$ 
in the physical domain of scalar curvatures (see \cite{Sotiriou:2008rp, DeFelice:2010aj, Nojiri:2017ncd}). We show that many considerably simple $f (R)$ theories lead to potentials $V (\phi)$ having a `hilltop' shape in the domain $\phi > 0$, exponentially decreasing to zero as $\phi \to \infty$.  The hilltop, depending on the parameters of the theory, can be more or less extended in 
$\phi$-space, which results in a flattened hilltop, or {\em tabletop}. In the Starobinsky model, 
this tabletop acquires the form of a plateau which extends to infinity. Inflation in such theories is accompanied by the quantum diffusion of the scalaron in the plateau region, and a universe of our type emerges in spatial regions in which the scalaron eventually rolls down to the stable minimum at $\phi = 0$ (corresponding to $R = 0$ in the Jordan frame), where it reheats the universe.  The spatial regions that end up in the asymptotic domain $\phi \to \infty$ run either into a `Big-Rip' singularity, or to an eternally inflating
universe. Eternal inflation is typical of all models in which $f (R)$ diverges at a 
finite value of the scalar curvature $R = R_m$, while Big-Rip singularity is characteristic of models with $f (R) \propto R^{1+\alpha}$ with $\alpha > 1$ as $R \to \infty$.  All such models exhibit asymptotically vanishing effective gravitational coupling, $G_{\rm eff} = G / f' (R) \to 0$, in the Jordan frame in the corresponding limit, implying vanishing gravitational interaction of matter and gravitons (see Appendix~\ref{sec:scatter}).

Models with simple hilltop scalaron potentials appear in those $f (R)$ theories that have a single mass scale, while flattened hilltops (tabletops) appear in models containing several different mass scales, such as $f (R) = R + R^2/m^2 + R^3/m_1^4$ with $m_1 \gg m$. In terms of inflationary predictions for the primordial power spectrum, tabletop models agree well with the Cosmic Microwave Background (CMB), while simple hilltops generally fail to reproduce the CMB power spectrum.  Thus, from the inflationary perspective, the higher-power corrections to the Starobinsky model should be suppressed in order that the predictions of the theory agree with observations. Regarding $f (R)$ gravity as a quantum correction to the general relativity theory, this probably gives us information about the underlying high-energy theory that produces such corrections.\footnote{This is equally valid for non-local extensions of $f (R)$ gravity in which a massive scalaron is the only extra degree of freedom \cite{Koshelev:2016xqb, Koshelev:2022olc}.}

Our paper is organised as follows. In Sec.~\ref{sec:direct}, we review the description of $f (R)$ gravity in the Jordan and Einstein frames. In Sec.~\ref{sec:poten}, we describe the general properties of the scalaron potential of $f (R)$ gravity, including its hilltop and tabletop features.  In Sec.~\ref{sec:free}, we describe a class of $f (R)$ gravity theories with limiting curvature in the Jordan frame; they exhibit asymptotically vanishing gravity as this limiting curvature is approached. Initial conditions for an inflationary universe are briefly discussed in Sec.~\ref{sec:IC}. In Sec.~\ref{sec:CMB_VE}, we study the inflationary evolution towards the stable minimum at $\phi = 0$ and determine the parameters characterising the primordial power spectrum for various models of interest. In Sec.~\ref{sec:asymptote}, we study the evolution in the region $\phi \to \infty$ of asymptotically vanishing gravity, showing that it runs either into an eternal inflationary regime or into a Big-Rip singularity depending on the $f (R)$ theory. We summarise our results in Sec.~\ref{sec:summary}. In Appendix~\ref{sec:scatter}, we establish the relation between two-particle scattering amplitudes in the Jordan and Einstein frames and show that two-graviton scattering cross-section tends to zero in the Jordan frame in the asymptotically vanishing gravity region. In Appendix~\ref{sec:odd}, we construct the scalaron potentials for theories where $f (R)$ is an odd function of $R$.

\section{Jordan and Einstein frames}
\label{sec:direct}

In this paper, we use the metric signature convention $(-, +, +, +)$. The action of the modified gravity theory under consideration is written as
\begin{equation} \label{Sg}
S_g = \frac{M_p^2}{3} \int d^4 x \sqrt{-g}\, f (R) \, ,
\end{equation}
where 
\begin{equation} \label{Pmass}
M_p = \sqrt{\frac{3}{16 \pi G}} 
\end{equation}
is a reduced Planck mass,\footnote{The Planck mass normalization (\ref{Pmass}) is convenient for the subsequent treatment of the theory in the Einstein frame, because the scalaron potential then depends on a simple expression $e^{\phi / M_p}$ [see \eqref{phi} below].} with $G$ being Newton's gravitational constant, so that, numerically, $M_p \approx 3 \times 10^{18}\, \text{GeV}$ in units $\hbar = c = 1$. In general relativity, $f_\text{GR} (R) = R - 2 \Lambda$, where $\Lambda$ is the cosmological constant.  It is customary to have the gravitational coupling constant present explicitly in front of the gravitational action (\ref{Sg}). 

The gravitational action (\ref{Sg}) describes gravity in the original so-called {\em Jordan\/} frame of field variables. Variation of the action with respect to the metric results in the gravitational equations
\begin{equation}\label{J}
f' (R) R_{\mu\nu} - \frac12 f (R) g_{\mu\nu} - \left( \nabla_\mu \nabla_\nu - g_{\mu\nu} \Box \right) f' (R) = 8\pi G\, T_{\mu\nu} \, , 
\end{equation}
which can be written as
\begin{equation}
R_{\mu\nu} - \frac{f (R)}{2 f' (R)} g_{\mu\nu} - \frac{\left( \nabla_\mu \nabla_\nu - g_{\mu\nu} \Box \right) f' (R)}{f' (R)} = 8\pi G_{\rm eff}\,T_{\mu\nu} \, ,
\label{mod_GR}
\end{equation}
where $G_{\rm eff} = G/f'(R)$.
Equation (\ref{mod_GR}) reduces to the familiar general-relativistic form
\begin{equation}
R_{\mu\nu} - \frac12 R g_{\mu\nu} = {8\pi G}\,T_{\mu\nu}
\end{equation}
when $ f = R $.
Equation (\ref{mod_GR}) serves to illustrate an important point, namely, that, in modified gravity theories (MOG), the gravitational coupling to matter changes from $G$ to 
$G_{\rm eff} = G/f'(R)$.
This could have important consequences especially if $f' \to \infty$ as 
$R \to R_m$. In this case, gravity becomes asymptotically vanishing as $R \to R_m$ since $G_{\rm eff} \to 0$. Compelling MOG theories 
with the property that $f/f' \to 0$ and $f' \to \infty$ are discussed later in this paper in Sec.~\ref{sec:free}.

The left-hand side of (\ref{J}) looks rather complicated 
since the metric obeys partial differential equations of the fourth order and,
therefore, contains one extra degree of freedom compared to general relativity. 
For this reason, it is often more convenient to proceed to the field frame in which 
(\ref{J}) acquires the Einstein form, with
 the additional degree of freedom materialising as a scalar field. 

Proceeding to the {\em Einstein\/} frame, we first write action (\ref{Sg}) in the form
\begin{equation}\label{Sg1}
S_g = \frac{M_p^2}{3} \int d^4 x \sqrt{-g}\, \bigl[ \Omega R - h (\Omega) \bigr] \, ,
\end{equation}
where $\Omega$ is a new dimensionless field, and $h (\Omega)$ is a Legendre transform of $f (R)$.  It is defined by the equations
\begin{align} \label{dir1}
f' (R) &= \Omega \quad \Rightarrow \quad R = R (\Omega) \, , \\
h (\Omega) &= \bigl[ \Omega R - f (R) \bigr]_{R = R (\Omega)} \, . \label{dir2}
\end{align}
The inverse transform allows one to find $f (R)$ given $h (\Omega)$, and is obtained by the variation of (\ref{Sg1}) with respect to $\Omega$:
\begin{align} \label{inv1}
h' (\Omega) &= R \quad \Rightarrow \quad \Omega = \Omega (R) \, , \\
f (R) &= \bigl[ \Omega R - h (\Omega) \bigr]_{\Omega = \Omega (R)} \, . \label{inv2}
\end{align}

These transformations may involve subtleties as to which solution is to be chosen in (\ref{dir1}) and (\ref{inv1}). The solution of these equations is unique for convex functions, e.g., if $f'' (R) > 0$ everywhere in the domain of validity. This is the stability condition for $f (R)$ gravity, which we mostly assume to be satisfied. Another stability condition is $f' (R) > 0$, which ensures the effective gravitational coupling to be positive in the Jordan frame, as is clear from (\ref{mod_GR}). This condition then implies $\Omega > 0$. For reviews of these general properties of $f (R)$ gravity, see \cite{Sotiriou:2008rp, DeFelice:2010aj, Nojiri:2017ncd}.   

As a next step, one transforms the action (\ref{Sg1}) so that the term which is linear in the scalar curvature takes the Einstein form. For this purpose, one performs a conformal transformation of the metric
\begin{equation} \label{conform}
g_{\mu \nu} = \Omega^{-1} \tilde g_{\mu\nu} \, .
\end{equation}
With this transformation, one gets
\begin{equation}
\frac{M_p^2}{3} \sqrt{- g}\, \Omega R = \frac{M_p^3}{3} \sqrt{- \tilde g}\, \left[ \tilde R - \frac32 \left( \tilde \nabla \ln \Omega \right)^2 + 3\, \tilde \Box \ln \Omega \right] \, ,
\end{equation}
in which all new metric-related objects are denoted by tildes. The last term is the total derivative and can be dropped. The transformed action (\ref{Sg1}) then becomes
\begin{equation}\label{Sg2}
S_g = \int d^4 x \sqrt{- \tilde g}\, \left[ \frac{M_p^2}{3} \tilde R - \frac{M_p^2}{2} \left( \tilde \nabla \ln \Omega \right)^2 - W (\Omega) \right] \, ,
\end{equation}
where
\begin{equation}\label{W}
W (\Omega) = \frac{M_p^2}{3} \frac{h (\Omega)}{\Omega^2} \, .
\end{equation}

We have thus obtained an Einstein theory of gravity with a minimally coupled scalar field.  To bring it to a canonical form, we introduce a scalar field (scalaron) $\phi$ by setting
\begin{equation} \label{phi}
\Omega = e^{\phi / M_p} \, .
\end{equation} 
Action (\ref{Sg2}) eventually becomes
\begin{equation}\label{Sg3}
S_g =  \int d^4 x \sqrt{- \tilde g}\, \left[ \frac{M_p^2}{3} \tilde R - \frac12 \left( \tilde \nabla \phi \right)^2 - V (\phi) \right] \, ,
\end{equation}
where the scalaron potential $V (\phi)$ is calculated by using (\ref{dir2}), (\ref{W}) and (\ref{phi}):
\begin{equation} \label{V}
V (\phi) \equiv W \left( \Omega (\phi ) \right) = \frac{M_p^2}{3} \left[ \frac{R}{\Omega} - \frac{f (R)}{\Omega^2} \right]_{\substack{R = R (\Omega) \\ \Omega = \Omega (\phi)}} \, .
\end{equation}

The relations (\ref{inv1}), (\ref{inv2}) give a solution of the inverse problem of finding $f (R)$ if $V (\phi)$ in the Einstein frame is known by constructing $h (\Omega)$ from (\ref{W}). Both direct and inverse problems may have subtleties as to which root of the corresponding equations (\ref{dir1}) and (\ref{inv1}) one should pick.  Several branches of solutions may arise when proceeding from the Jordan frame to the Einstein frame (see below).

Using relations (\ref{dir1}) and (\ref{dir2}), it is easy to establish that the scalaron potential will have an extremum, with $V' (\phi) = 0$, at the Jordan-frame value of
 $R$ which satisfies
\begin{equation} \label{statR}
R f' (R) = 2 f (R) \, .
\end{equation}
The scalaron mass squared, $m_\phi^2 = V'' (\phi)$, at this extremum is given by
\begin{equation}\label{mphi}
m_\phi^2 = \frac13 \left[ \frac{1}{f'' (R)} - \frac{R}{f' (R)} \right] = \frac13 \left[ \frac{1}{f'' (R)} - \frac{R^2}{2 f (R)} \right] \, .
\end{equation}
If $m_\phi^2 > 0$, then this is a local minimum.  Note that the scalaron potential as a function of $R$ is given by [see (\ref{V})]
\begin{equation} \label{U}
U (R) \equiv V \left( \phi (R) \right) = \frac{M_p^2}{3} \left[\frac{R}{f' (R)} - \frac{f (R)}{\left( f' (R) \right)^2}\right]  \, .
\end{equation}
This, in particular, implies that proceeding from Lagrangian $f (R)$ to $ - f (- R)$ just produces an inverted potential $V (\phi) \to - V (\phi)$.  

The stress-energy tensor of matter in the Einstein frame is related to that in the Jordan frame by
\begin{equation} \label{T}
\tilde T_{\mu\nu} = \Omega^{-1} T_{\mu\nu} = e^{- \phi / M_p} T_{\mu\nu} \, .
\end{equation}

\section{General properties of the scalaron potential}
\label{sec:poten}

In this section, we establish the typical qualitative behaviour of the scalaron potential (\ref{V}).  If we assume that the function $f (R)$ has a general analytic behaviour in the neighbourhood of $R = 0$, then we can expand it in powers of $R$:
\begin{equation} \label{fstar}
f (R) = - 2 \Lambda + R + \frac{R^2}{m^2} + \ldots \, ,
\end{equation}
where $\Lambda$ is the cosmological constant, and $m$ is a parameter of dimension mass.  With 
the cosmological constant being small, one can see from \eqref{statR} and \eqref{mphi} that the theory has a stable minimum at $R \approx 4 \Lambda$ with the scalaron mass squared $m_\phi^2 = m^2 / 6 + {\cal O} ( \Lambda) $. Neglecting the cosmological constant and truncating (\ref{fstar}) at the term $R^2$, one obtains the famous Starobinsky model \cite{Starobinsky:1980te, Vilenkin:1985md} that was the basis of one of the first inflationary models of the universe. The scalaron potential (\ref{V}) in this case is \cite{Whitt:1984pd}
\begin{equation} \label{Vstar}
V (\phi) = 
\frac{M_p^2 m^2}{12} \left( 1 - e^{- \phi / M_p} \right)^2 \, ,
\end{equation}
so that
\begin{equation}
V (\phi) \approx \frac{1}{12} m^2 \phi^2 \, , ~~ {\rm for} ~~ \phi \ll M_p \, ,
\end{equation}
where $m \approx \sqrt{6}\times 10^{-5}\,M_p$ is set by CMB observations.
All modified gravity models of the form (\ref{fstar}) possessing a non-vanishing
 $R^2$ term in $f(R)$ will lead to 
potentials having a similar quadratic behaviour near the minimum of $V(\phi)$. This feature of the potential is of importance since, after the end of inflation, the scalaron will oscillate around $\phi=0$ thereby reheating the universe in the conventional manner (i.e., either through perturbative reheating or via non-perturbative resonance-like preheating).

Let us now consider the asymptotic properties of the scalaron potential (\ref{V}) as $\phi \to \infty$. Since we have $f'' (R) > 0$ by assumption, the derivative $f' (R)$ is growing with $R$.  Assuming that it grows unbounded as $R \to \infty$, by virtue of (\ref{dir1}) and (\ref{phi}), the functions $\Omega (R)$ and $\phi (R)$ will also be growing and tending to infinity as $R \to \infty$.  The asymptotic behaviour of the scalaron potential will then be determined by the behaviour of expression (\ref{U}) in the limit of large $R$.

\begin{enumerate}
\item
As a first example, consider the case $f (R) \sim R^{1 + \alpha} / m^{2\alpha}$ at large $R$, where $m$ is a mass parameter needed for dimensional reasons.  The stability condition $f'' (R) > 0$ requires $\alpha > 0$.  The potential (\ref{U}) in this case asymptotically behaves as
\begin{equation} \label{ualpha}
U (R) \sim \frac{M_p^2 m^2}{3} \frac{\alpha}{(1 + \alpha)^2} \left( \frac{R}{m^2} \right)^{1 - \alpha} \, .
\end{equation}
In terms of the scalaron $\phi$, one obtains
\begin{equation}\label{alpha}
V (\phi) \sim \frac{M_p^2 m^2}{3} \frac{\alpha}{(1 + \alpha)^{1+1/\alpha}} \exp \left[ \left( \frac{1}{\alpha} - 1 \right) \frac{\phi}{M_p} \right] \, .
\end{equation}
For $\alpha = 1$ (which includes the Starobinsky model), the scalaron potential is asymptotically flat; for $\alpha < 1$, it is growing, while for $\alpha > 1$, it asymptotically declines to zero.  Such potentials have been studied in \cite{Carloni:2004kp, Ivanov:2011np, Bukzhalev:2013saa} and 
other papers. 

{\em From these relations, it is clear that MOG models with Lagrangians $f (R)$ which are analytic at $R = 0$ and grow faster than $R^2$ as $R \to \infty$ will give rise to `hilltop' scalaron potentials in the domain $\phi \geq 0$.}  

\item
As a concrete example from the above category, consider the MOG model 

\begin{equation} \label{pw}
f (R) = R \left[ 1 + \frac{1}{n + 1} \left( \frac{R}{m^2} \right)^n\, \right]
\end{equation}
for positive $n$. 
Equation \eqref{dir1} in this case reads
\begin{equation} \label{eqr-pw}
1 + \left( \frac{R}{m^2} \right)^n = \Omega \, .
\end{equation}
Its solution in the domain $R > 0$ is given by
\begin{equation} \label{Rodd}
R (\Omega) = m^2 \left( \Omega - 1 \right)^{1/n} \, ,
\end{equation}
and the potential (\ref{W}) in this region $(\Omega > 1)$ is
\begin{equation} \label{wodd}
W (\Omega) = \frac{M_p^2}{3} \left[\frac{R (\Omega)}{\Omega} - \frac{f\left( R (\Omega) \right)}{\Omega^2}\right] 
= \frac{M_p^2 m^2}{3} \left(1+ \frac{1}{n} \right) \frac{\left( \Omega - 1 \right)^{1+ 1/n}}{\Omega^2} \, .
\end{equation}
From \eqref{Rodd} and \eqref{wodd}, one can see that the theory actually makes sense for all signs of $R$ and values of $\Omega$ if either $n$ or $1/n$ is an odd integer. For even integer $n$, the function $f (R)$ becomes odd, and one can extend the solution to the region $\Omega < 1$ as described in Appendix~\ref{sec:odd}.  The scalaron potential is then found to be
\begin{equation} \label{pwpot}
V (\phi) = \dfrac{M_p^2 m^2}{3} \left(1+ \frac{1}{n} \right) \times \left\{ \begin{array}{ll} 
\left( e^{\phi / M_p} - 1 \right)^{1+ 1/n} e^{- 2 \phi / M_p} \, , \quad \text{odd $n$ or $\dfrac{1}{n}$} \, , \bigskip \\ \text{sign} \left( \phi \right)
\left( e^{|\phi| / M_p} - 1 \right)^{1+ 1/n} e^{- 2 |\phi| / M_p} \, , \quad \text{even $n$} \, .
\end{array} \right.
\end{equation}
For large values of $\phi > 0$, one finds 
\begin{equation}
V(\phi) \propto \exp{\left[\left (\frac{1}{n} - 1\right )\frac{\phi}{M_p}\right]} \, .
\end{equation}
Therefore, for $n > 1$, $V(\phi)$ declines to zero as $\phi \to \infty$ resulting in a hilltop potential. For $n<1$, on the other hand, $V(\phi)$ grows exponentially with $\phi$ and can give rise to power law inflation which is ruled out by CMB observations
\cite{Planck:2018jri,PLI}. The value $n=1$ implies that at large values of $\phi$ 
the potential is described by a plateau with $V(\phi) \to {\rm constant}$,
which is a key feature of the Starobinsky inflation (\ref{Vstar}).

In the neighbourhood of $\phi = 0$, the potential (\ref{pwpot}) has the form
$V (\phi) \propto \phi^{1 + 1/n}$ for odd $n$ or $1/n$. It is well known that the averaged equation of state (EOS) during oscillations of the scalar field about the minimum of a potential having the form
$V(\phi) \propto \phi^{2p}$ is \cite{Turner:1983he}
\begin{equation}
\langle w_\phi\rangle = \frac{p-1}{p+1}~.
\end{equation}
Substituting $p = \frac{1}{2}\left (1+\frac{1}{n}\right )$, one gets
\begin{equation}
\langle w_\phi\rangle = \frac{1-n}{1+3n} \, ,
\end{equation}
implying $\langle w_\phi\rangle = 0$ for $n=1$ and $-1/3 < \langle w_\phi\rangle < 0$ for $n>1$. In other words, the EOS of the oscillating scalar can become negative for large values of $n$ which makes it prone to the onset of instabilities, as noted in \cite{Johnson:2008se}. For $n < 1$, the reverse is true since $\langle w_\phi \rangle > 0$. In this case, the EOS of the oscillating scalar is positive and can become `radiation-like' with  $\langle w_\phi \rangle = 1/3$ for $n = 1/3$.

\item Our third example is provided by the MOG model
\begin{equation} \label{fexpon}
f (R) = R \, e^{R / m^2} \, .
\end{equation}
The function $f (R)$ now grows faster than any power of $R$. Equation (\ref{dir1}) in this case reads
\begin{equation} 
\left( 1 + \frac{R}{m^2} \right) e^{R/m^2} = \Omega \, .
\end{equation}
The function on the left-hand side is monotonic in the domain where it is positive, and the solution $R (\Omega)$ is unique. It is given by
\begin{equation}
R (\Omega) = m^2 \left[ w ( e \Omega ) - 1 \right] \, ,
\end{equation}
where $w (x)$ is the Lambert function, which is a solution of the equation $x = w e^w$. 
The scalaron potential is determined to be
\begin{equation} \label{Vexpon}
V (\phi) = \frac{M_p^2 m^2}{3} \left. \frac{\left[ w(e \Omega) - 1 \right]^2}{\Omega w (e \Omega)} \right|_{\Omega = e^{\phi / M_p}} = \frac{M_p^2 m^2}{3}  \frac{\left[ w \left( e^{\phi / M_p + 1} \right) - 1 \right]^2}{w \left( e^{\phi / M_p + 1} \right)} e^{- \phi / M_p} \, ,
\end{equation}
and is plotted in Fig.~\ref{fig:expon} together with potential (\ref{Vstar}) of the Starobinsky model.  Asymptotically, it behaves as
\begin{equation} \label{exponent}
V (\phi) \approx \frac{M_p^2 m^2}{3} \left( \frac{\phi}{M_p} - \ln \frac{\phi}{M_p} \right) e^{- \phi / M_p} \, , \qquad \frac{\phi}{M_p} \gg 1 \, .
\end{equation}

\begin{figure}[ht]
\begin{center}
\includegraphics[width=.66\textwidth]{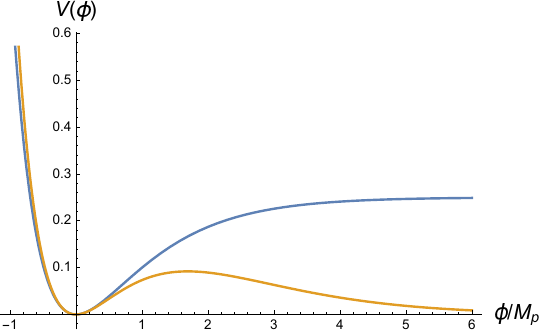}
\caption{Potential (\ref{Vstar}) of the Starobinsky model (blue) and the `hilltop' potential (\ref{Vexpon}) (orange) plotted in units $M_p^2 m^2 / 3$. \label{fig:expon}}
\end{center}
\end{figure}

In the opposite limit $\phi \to - \infty$, or $\Omega \to 0$, the scalaron potential typically grows unbounded.  Indeed, according to (\ref{dir1}), this corresponds to $f' (R) \to 0$, and the second term in (\ref{U}) then dominates and tends to infinity.  In the Starobinsky model (\ref{fstar}), this takes place at $R = - m^2/2$, and in the exponential model (\ref{fexpon}) at $R = - m^2$.  In both these cases, the potential steeply grows as $V (\phi) \propto e^{-2 \phi / M_p}$ for large negative values of $\phi$.  
This can also be observed in Fig.~\ref{fig:expon}.

Generalising \eqref{fexpon}, suppose the function $f (R)$ has an asymptotic form
\begin{equation} \label{fexpow}
f (R) \propto R^\alpha e^{\left(R  /m^2\right)^\beta}
\end{equation}
as $R \to \infty$, with $\alpha > 0$ and $\beta > 0$. In this case, we will have, asymptotically,
\begin{equation} 
f '(R) = e^{\phi / M_p} \propto R^{\alpha + \beta - 1} e^{\left(R  /m^2\right)^\beta} \, , \qquad
R \propto \phi^{1/\beta} \, .
\end{equation}
The first term in the potential \eqref{V} will dominate at large $\phi$, and we will have
\begin{equation} \label{Vexpow}
V(\phi) \propto R (\phi) e^{- \phi  /M_p} \propto \phi^{1/\beta} e^{- \phi  /M_p} \, .
\end{equation}
Asymptotic evolution for potentials of this form will be considered in Sec.~\ref{sec:inflation}.

\item{\em MOG models containing several scales}

MOG models with several mass scales have the attractive feature of flattening the hilltop and producing an extended plateau which we refer to as a tabletop.  

\begin{figure}[ht]
\begin{center}
\includegraphics[width=.7\textwidth]{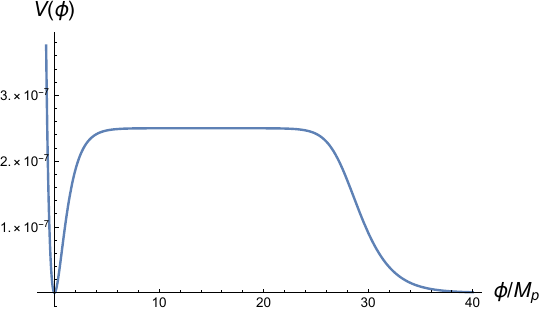}
\caption{The tabletop  potential corresponding to Lagrangian (\ref{cubic}) with $m_1 = M_p$ and $m = 10^{-3} M_p$ in units $M_p^4/3$. Note that values of $\phi$ close to the hilltop can result in inflation. However if  $\phi$ subsequently rolls towards $\phi \to \infty$ then the universe will encounter a future `Big-Rip' singularity in the Jordan frame, as shown in Sec.~\ref{sec:asym}.
\label{fig:scales}}
\end{center}
\end{figure}

Consider, for instance, the MOG model obtained by adding a cubic term to
 (\ref{fstar}):
\begin{equation}\label{cubic}
f (R) = R + \frac{R^2}{m^2} + \frac{R^3}{m_1^4} \, ,
\end{equation}
with $m_1 \gg m$.  The scalaron potential can be calculated analytically:
\begin{equation} \label{cubicpot}
V (\phi ) = \frac{2 M_p^2 m_1^8}{81 m^6} e^{-2 \phi / M_p} s^2 (\phi) \left[ s (\phi) + \frac{3}{2} \right] \, , 
\end{equation}
where
\begin{equation}
s (\phi) = \sqrt{ 1 + \frac{3 m^4}{m_1^4} \left( e^{\phi/M_p} - 1 \right)} - 1 \, ,
\end{equation}
and its shape is shown in Fig.~\ref{fig:scales} (in units $M_p^4/3$) for $m_1 = M_p$ and $m = 10^{-3} M_p$. For smaller values of $m$, the tabletop will be more extended.  In fact, the left-hand slope of the plateau is similar to that of the Starobinsky model, while the right-hand slope decreases asymptotically as 
\begin{equation}\label{ascub}
V (\phi) \approx \frac{2 M_p^4}{9 \sqrt{3}} e^{- \phi / 2 M_p} \,.
\end{equation}
We will see in Sec.~\ref{sec:asym} that, if the inflaton rolls towards the right, then  this model will ultimately encounter a future `Big-Rip'  singularity in the Jordan frame. 

\end{enumerate}

\section{Asymptotically vanishing gravity}
\label{sec:free}

We have already mentioned that, as $f' (R)$ tends to infinity (either at finite or at infinite curvature $R$), the effective gravitational coupling $G_\text{eff} = G / f' (R)$ vanishes; see (\ref{mod_GR}). This is also clear from (\ref{conform}) and (\ref{T}), with the latter demonstrating that the energy density and pressure of matter acquire the factor $\Omega^{-2} = e^{- 2\phi / M_p}$ in the Einstein frame, vanishing in the limit $\phi \to \infty$, corresponding to $f' (R) \to \infty$ by virtue of (\ref{dir1}).  Gravitational interaction vanishes in this limit not only for matter but also for gravitons. In particular, as we show in Appendix~\ref{sec:scatter}, the two-graviton scattering cross-section with fixed collision energy $\cE$ in the Jordan frame behaves as $\sigma \sim G_\text{eff}^2\, \cE^2$, vanishing in the limit $G_\text{eff} \to 0$.

An interesting special class of MOG models are those for which both $f (R)$ and $f' (R)$ become infinite at a {\em finite\/} value of $R = R_m > 0$. The scalar curvature $R_m$ is then the maximal scalar curvature in the Jordan frame. We assume a naturalness condition
\begin{equation}
\lim_{R \to R_m} \frac{f (R)}{f'(R)} = 0 \, .
\end{equation}
In this case, the scalaron potential (\ref{V}) has the asymptotic form
\begin{equation}\label{scalexp}
V (\phi) \approx \frac{M_p^2 R_m}{3 \Omega (\phi)} = \frac{M_p^2 R_m}{3} e^{- \phi / M_p} \, , \qquad \frac{\phi}{M_p} \gg 1 \, .
\end{equation}
As will be shown in Sec.~\ref{sec:inflation}, scalaron potentials with the asymptotic behaviour \eqref{scalexp} 
will result in eternal inflation in the region $\phi \gg M_p$. Let us give some examples.

\begin{itemize}

\item
Our first example is the MOG model
\begin{equation} \label{power}
f (R) = \frac{R}{\left( 1 - R / R_m \right)^{\beta}} \, , \qquad f' (R) = \frac{1 + (\beta - 1) R / R_m}{\left( 1 - R / R_m \right)^{\beta + 1} } \, ,
\end{equation}
with $\beta > 0$.  Equation (\ref{dir1}) then gives, asymptotically,
\begin{equation}
\frac{\beta }{\left( 1 - R / R_m \right)^{\beta + 1} } = \Omega \impl R = R_m \left[ 1 - \left( \frac{\beta}{\Omega} \right)^{1/(\beta + 1)} \right] \, .
\end{equation}
The scalaron potential is calculated to be
\begin{equation}
V (\phi) \simeq \frac{M_p^2 R_m}{3 \Omega (\phi)} \left[ 1 - \frac{\beta + 1}{\beta} \left( \frac{\beta}{\Omega (\phi)} \right)^{1/(\beta + 1)} \right] 
\, ,
\end{equation}
reproducing (\ref{scalexp}) in the asymptotic region $\Omega (\phi) = e^{\phi/M_p} \gg \beta $.

As a particular case, consider model (\ref{power}) with $\beta = 1$. Equation (\ref{dir1}) in the domain $R < R_m$ has the solution
\begin{equation}
R (\Omega) = R_m \left( 1 - \frac{1}{\sqrt{\Omega}} \right) \, ,
\end{equation}
with the corresponding scalaron potential resembling a hilltop
\begin{equation} 
V (\phi) = \frac{M_p^2 R_m}{3} \frac{\left( \sqrt{\Omega (\phi)} - 1 \right)^2}{\Omega^2 (\phi) } = \frac{M_p^2 R_m}{3} e^{- \phi/M_p} \left( 1 - e^{-\phi/ 2 M_p} \right)^2 \, ,
\label{eq:scalaron_pole}
\end{equation}
which is shown in Fig.~\ref{fig:ration} in blue.
For large values of $\phi \gg M_p$ this potential has the asymptotic form
$V (\phi) \propto e^{- \phi/M_p}$ whose significance will be discussed later in
Sec.~\ref{sec:asymptote}.

\begin{figure}[ht]
\begin{center}
\includegraphics[width=.7\textwidth]{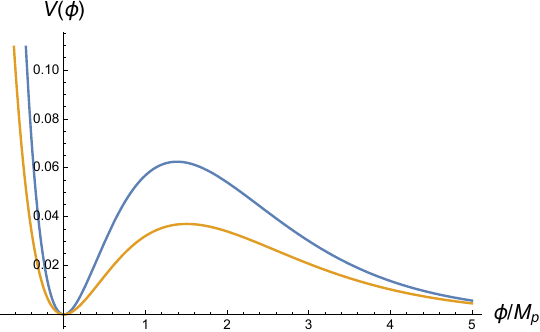}
\caption{
Scalaron potentials corresponding to theory (\ref{power}) with $\beta = 1$ ({\em blue\/}) and $\beta = 2$ ({\em orange\/}) in units $M_p^2 R_m / 3$. \label{fig:ration}}
\end{center}
\end{figure}

The MOG model (\ref{power}) with $\beta = 2$ is also solvable analytically, but the expression for the scalaron potential is rather complicated. The corresponding potential is shown in Fig.~\ref{fig:ration} in orange.

\item An analytically solvable example of asymptotically vanishing gravity with two mass scales is provided by the rational function
\begin{equation}\label{ration-scales}
f (R) = R\, \frac{1 + R / m^2}{1 - R/m_1^2} \, ,
\end{equation}
If $m_1 \gg m$, then this model reproduces the Starobinsky model for $|R| \ll m_1^2$. 
It also has a pole at $R = m_1^2$.  The scalaron potential for (\ref{ration-scales}) is given by
\begin{equation} \label{Vscales}
V (\phi) = \frac{M_p^2 m_1^2}{3} e^{- 2 \phi / M_p} \left( \sqrt{ e^{\phi / M_p} + \frac{m_1^2}{m^2}} - \sqrt{ 1 + \frac{m_1^2}{m^2}} \right)^2 \, .
\end{equation}
$V(\phi)$ interpolates between the Starobinsky-model potential when $m_1 \to \infty$, 
and the potential (\ref{eq:scalaron_pole}) 
with $R_m = m_1^2$ when $m \to \infty$.  In fact, (\ref{Vscales})
 has a typical `hilltop' form already at $m_1 \simeq m$.

For $m_1 \gg m$, at intermediate values
\begin{equation}
1 \ll e^{\phi / M_p} \ll \frac{m_1^2}{m^2} \, ,
\end{equation}
the potential resembles a tabletop 
of height $V \approx M_p^2 m^2 / 12$.  The right-hand slope 
of the tabletop decreases asymptotically in accordance with (\ref{scalexp}):
\begin{equation}
V (\phi) \approx \frac{M_p^2 m_1^2}{3} e^{- \phi / M_p} \, , \qquad e^{\phi / M_p} \gg \frac{m_1^2}{m^2} \, .
\end{equation}

The scalaron potential (\ref{Vscales}) is shown in Fig.~\ref{fig:scales-rat} (in units $M_p^4/3$) for $m_1 = M_p$ and $m = 10^{-3} M_p$.  For smaller values of $m$, the tabletop region is more extended.

\begin{figure}[ht]
\begin{center}
\includegraphics[width=.7\textwidth]{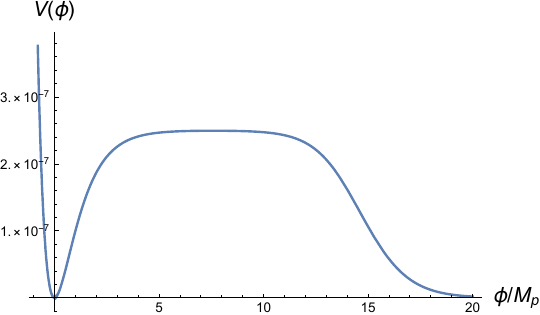}
\caption{The tabletop potential corresponding to the MOG Lagrangian (\ref{ration-scales}),
namely (\ref{Vscales})
 with $m_1 = M_p$ and $m = 10^{-3} M_p$ in units $M_p^4/3$. 
Note that values of $\phi$ close to the hilltop
will result in inflation. An inflaton moving towards the left, to smaller
values of $\phi$, will subsequently oscillate around $\phi = 0$ resulting
in post-inflationary reheating. However if the inflaton
rolls towards the right then the universe
will begin to inflate eternally in the Jordan frame when $\phi \to \infty$
(this is further discussed in Sec.~\ref{sec:asym}).
Therefore, two dramatically different future possibilities are permitted by
potentials of the type (\ref{Vscales}).
\label{fig:scales-rat}}
\end{center}
\end{figure}

\end{itemize}

\section{Initial conditions for an inflationary universe}
\label{sec:IC}

We have shown that
hilltop scalaron potentials (with or without an extended plateau), are
 generic in inflationary models based on $f (R)$ gravity. The
question of initial conditions naturally arises for this class of models.  A common wisdom is that such a universe could be created in the region $\phi > 0$ in the Einstein frame.  Assuming this, and adopting the tunnelling boundary conditions \cite{Vilenkin:1983xq, Linde:1983cm, Linde:1983mx, Zeldovich:1984vk, Rubakov:1984ki, Rubakov:1984bh, Vilenkin:1984wp, Vilenkin:1985md, Vilenkin:1986cy}, we will have, for the logarithm of the probability of creation,
\begin{equation}
\ln \cP_\text{c} (\phi) \approx - \frac{32 \pi^2 M_p^4}{3 V (\phi)} \, . 
\label{eq:quant_creation}
\end{equation}
In this case, the universe is most likely to be created with the largest possible value of the potential; hence, in the inflationary region $\phi > 0$, the scalaron initially will most likely be at the hilltop of its potential. 

During quantum tunnelling, the universe in the Einstein frame is created spatially closed with the initial scale factor (size)
\begin{equation} \label{size}
\tilde a (\phi) \simeq M_p \sqrt{ \frac{2}{V (\phi)}} \, . 
\end{equation}
The initial velocity of the scalaron in this approximation is distributed around zero with dispersion $\langle \dot \phi^2 \rangle \ll V (\phi)$.  For all practical purposes, 
when extrapolating the wave function to the creation point, one can assume 
initial conditions (\ref{size}) and $\dot \phi = 0$ (which jointly imply $H = 0$) at some initial value of $\phi$ close to (but away from) the maximum of the potential $V (\phi)$.  These initial conditions can lead to a period of slow-roll 
inflation, during which the spatial curvature of the universe quickly becomes unimportant.

Regardless of the specific theory of quantum creation of the universe, quantum limitations on the classical initial conditions for chaotic inflation in models with modified gravity were recently discussed in \cite{Gorbunov:2014ewa, Gorbunov:2016fxw}.  The idea of this analysis consists in considering an initial patch of the universe in the Jordan frame and determine the natural limiting energy-density scale $\rho$ in this frame.  For the $f(R)$ models  (\ref{Sg}), this gives 
\begin{equation}
\rho \sim M_p^2 f (R) \, .
\end{equation}
By virtue of (\ref{conform}) and (\ref{T}), the corresponding energy-density scale $\tilde \rho$ in the Einstein frame is given by 
\begin{equation}
\tilde \rho = \Omega^{-2} \rho \sim M_p^2 \frac{f (R)}{\left(f' (R) \right)^2} \sim U (R) = V (\phi) \, ,
\end{equation}
where $U (R)$ was defined in (\ref{U}).  Therefore, the limiting classical energy scale in the initial patch in the Einstein frame is comparable to the value of the scalaron potential, making the inflationary regime quite probable.  The spatial size $\tilde \ell$ of the initial classical patch in the Einstein frame is limited from below by
 \begin{equation}
\tilde \ell \gtrsim \tilde \rho^{\,- 1/4} \sim V^{- 1/4} (\phi) \, .
\end{equation}
Since typically one has $V (\phi) \ll M_p^4$, the size (\ref{size}) of a closed
quantum-mechanically created universe satisfies this constraint with a large margin.

During inflation, the scalar field experiences quantum diffusion \cite{Vilenkin:1983xq, Starobinsky:1986fx}.  If inflation commences somewhere in the region where the potential is plateau-like as shown in Fig.~\ref{fig:scales-rat}, then diffusion will proceed effectively in both directions, and a random hypothetical observer in such a universe will have equal chances of moving either towards the usual inflationary slope to the left with relaxation at the stable point $\phi = 0$, or towards the asymptotic region $\phi \to \infty$ to the right. Approximating the plateau by a flat potential, the respective chances will be proportional to the distance in the $\phi$-space from the scalar initial position on the plateau to the commencement of the slope in the two complementary directions.  From this perspective, the chances for a random observer to evolve to the right slope and to the left slope are practically equally high.  Note, however, that quantum diffusion being a branching process in the inflationary universe, the appearance of both type of regions in the universe as a whole is almost a certainty.

In this world picture, those regions of the universe that eventually diffuse to the left of the plateau resemble, after reheating, the universe which we observe. We will study this evolution in the next section.  In Sec.~\ref{sec:asymptote}, we will show that those regions that diffuse to the right end up either in an eternally inflating universe (as in the case of $f (R)$ with limiting curvature) or in a Big-Rip singularity.  

\section{Primordial power spectra from $f(R)$ inflation}
\label{sec:CMB_VE}

In this section, we examine the evolution of the scalar field as it
rolls down the left slope of the hilltop/tabletop. In this case
the universe ends up in a stable state around $\phi = 0$.  We are going to determine the parameters of the inflationary primordial power spectra for our models and make conclusions about their viability.  First, we shall consider simple hilltop models with a single mass scale $m$, and then models with two mass scales $m$ and $m_1$ whose scalaron potentials resemble flattened hilltops (tabletops).

Our numerical results describe a spatially closed universe with the initial conditions prescribed as indicated in the previous section after equation \eqref{size}. We consider the initial values of $\phi$ such that inflation lasts long enough to solve the flatness problem.

\bigskip

   We shall focus our attention to the homogeneous and isotropic FRW universe with the metric 
\begin{equation}
d \tilde s^{\,2} = - d \tilde t^{\,2} + \tilde a^2 ( \tilde t)\, d s_\kappa^2 \, ,
\end{equation}
where $d s_\kappa^2$ is the metric of the homogeneous isotropic three-space with curvature labelled by $\kappa = 0, \pm 1$.  We denote the metric variables in the Einstein frame by a tilde. For a given scalaron potential $V(\phi)$, the Einstein's equations lead to 
\begin{align}
\tilde{H}^2 &\equiv \l( \frac{\dot{\tilde a}}{\tilde a} \r)^2 = \frac{1}{2M_p^2} \, \rho_\phi - \frac{k}{\tilde{a}^2} ~, \label{eq:E_Hubble}\\
\frac{\ddot{\tilde a}}{\tilde a} &= -\frac{1}{4M_p^2} \l( \rho_\phi + 3 \, p_\phi \r) ~,\label{eq:E_acc}
\end{align}
where $\tilde a$ is the scale factor in the Einstein frame, and the overdot denotes the derivative with respect to the implicit time variable, in this case $\tilde t$.  The expressions for density $\rho_\phi$ and pressure $p_\phi$ are given by
\begin{align}
\rho_\phi &= \frac{1}{2}\dot{\phi}^2 + V(\phi) ~, \label{eq:E_rho}\\
p_\phi &= \frac{1}{2}\dot{\phi}^2 - V(\phi) ~, \label{eq:E_pr}
\end{align}
and the scalaron (inflaton) equation of motion is given by 
\begin{equation}
\ddot{\phi} + 3 \tilde{H} \dot{\phi} + V'(\phi) = 0~,
\label{eq:E_EOM}
\end{equation}
where $V'(\phi) = {\rm d}V / {\rm d}\phi$. We definite the Hubble slow-roll parameters $\epsilon_H$, $\eta_H$ as
\begin{equation}
\epsilon_H = -\frac{\dot{\tilde H}}{\tilde H^2}\, , \qquad \eta_H = - \frac{\ddot{\phi}}{\tilde{H}\dot{\phi}}\, ,
\label{eq:epsH_etaH}
\end{equation}
and the potential slow-roll parameters $\epsilon_{_V}$, $\eta_{_V}$ as 
\begin{equation}
\epsilon_{_V} = \frac{M_p^2}{3} \l(\frac{V'}{V}\r)^2\, , \qquad \eta_{_V}= \frac{2M_p^2}{3}   \l(\frac{V''}{V}\r) \, .
\label{eq:epsV_etaV}
\end{equation}
As described  in the previous section,  we  consider `typical' or `most-likely' initial conditions for inflation  set by quantum creation closer to the hilltop maximum,  hence the effects of a  spatial  curvature  on  the inflation dynamics   in the asymptotic future (corresponding to the Hubble-exit epoch of CMB scales) are negligible. The condition for inflation, $\ddot{\tilde a}/\tilde{a} > 0$, translates to $\epsilon_H < 1$. Under the slow-roll approximation $\epsilon_H, \eta_H \ll 1$, we have $\epsilon_H \simeq \epsilon_{_V}$ and $\eta_H \simeq \eta_{_V} -  \epsilon_{_V}$. The scalar and tensor power spectra under slow-roll approximation are given by
\begin{equation}
P_{_S} = \frac{3}{16\pi^2}  \l(\frac{\tilde H}{M_p}\r)^2 \, \frac{1}{\epsilon_H} \, , \qquad P_{_T} = \frac{3}{\pi^2}  \l( \frac{\tilde H}{M_p} \r)^2  \, ,
\label{eq:PS_PT}
\end{equation}
with the corresponding spectral indices 
\begin{equation}
 n_{_S} -1 =  2\, \eta_H - 4 \, \epsilon_H \, , \qquad  n_{_T} =  -2\, \epsilon_H \, .
\label{eq:nS_nT}
\end{equation}
The tensor-to-scalar ratio is given by 
\begin{equation}
r = 16 \, \epsilon_H \, .
\label{eq:r}
\end{equation}

Note that $ M_p^2 = 3 / 16 \pi G$, as mentioned earlier. It is customary to test  the predictions for $\lbrace n_{_S} , r \rbrace$ of a given potential $V(\phi)$  against the CMB data. In what follows, we will carry out the same for the scalaron potentials corresponding to the important MOG models discussed in this work. We note that for single mass scale  hilltop models,  the parameters $\lbrace n_{_S} , r \rbrace$ are rather insensitive to the mass scale $m$, which sets the overall normalisation of the power spectrum. For models with two mass scales $m$ and $m_1$, parameters $\lbrace n_{_S} , r \rbrace$ are sensitive to their ratio, namely $m_1/m$.

\subsection[Model \ $f(R) =  R \, e^{R/m^2}$]{Model \ $\boxed{ f(R) =  R \, e^{R/m^2} }$}
\label{sec:ReR}

For model (\ref{fexpon}), the scalaron potential is given by \eqref{Vexpon}. Using the properties of the Lambert function $w (x)$, we derive expressions for the scalaron potential and its first two derivatives to be
\begin{align}
V (\phi) &=  \f{m^2 M_p^2}{3} \, \l[ w + \f{1}{w} - 2 \r] \, e^{-\phi / M_p} ~,   \label{eq:V_E_ReR} \\[3pt]
V' (\phi) &= \f{m^2 M_p^2}{3} \, \f{1}{M_p} \, \l[ 3 - w - \f{2}{w} \r] \,  e^{-\phi / M_p} ~,   \label{eq:V'_E_ReR} \\[3pt]
V''(\phi) &=  \f{m^2 M_p^2}{3} \, \f{1}{M_p^2} \, \l[  \f{ 4 - w - 3 w^2 + w^3 }{ w \l( 1+w \r) } \r] \,  e^{-\phi / M_p} ~,
\label{eq:V''_E_ReR}
\end{align}
where $w \equiv w \left( e\Omega \right) = w \bigl( e^{\f{\phi}{M_p}+1} \bigr)$. Using these expressions, the slow-roll parameters can be written as
\begin{align}
\epsilon_{_V} (\phi)  &=  \f{1}{3} \, \l[ \f{3-w-\f{2}{w}}{w+\f{1}{w}-2} \r]^2 \, , \label{eq:eV_Rer} \\[3pt]
\eta_{_V} (\phi)  &= \f{2}{3} \,   \l[ \f{4-w-3w^2+w^3}{w \l(1+w\r)\l(w+\f{1}{w}-2\r)} \r] \, . \label{eq:etaV_ReR}
\end{align}

\begin{figure}[htb]
\centering
\subfigure[][]{
\includegraphics[width=0.6\textwidth]
{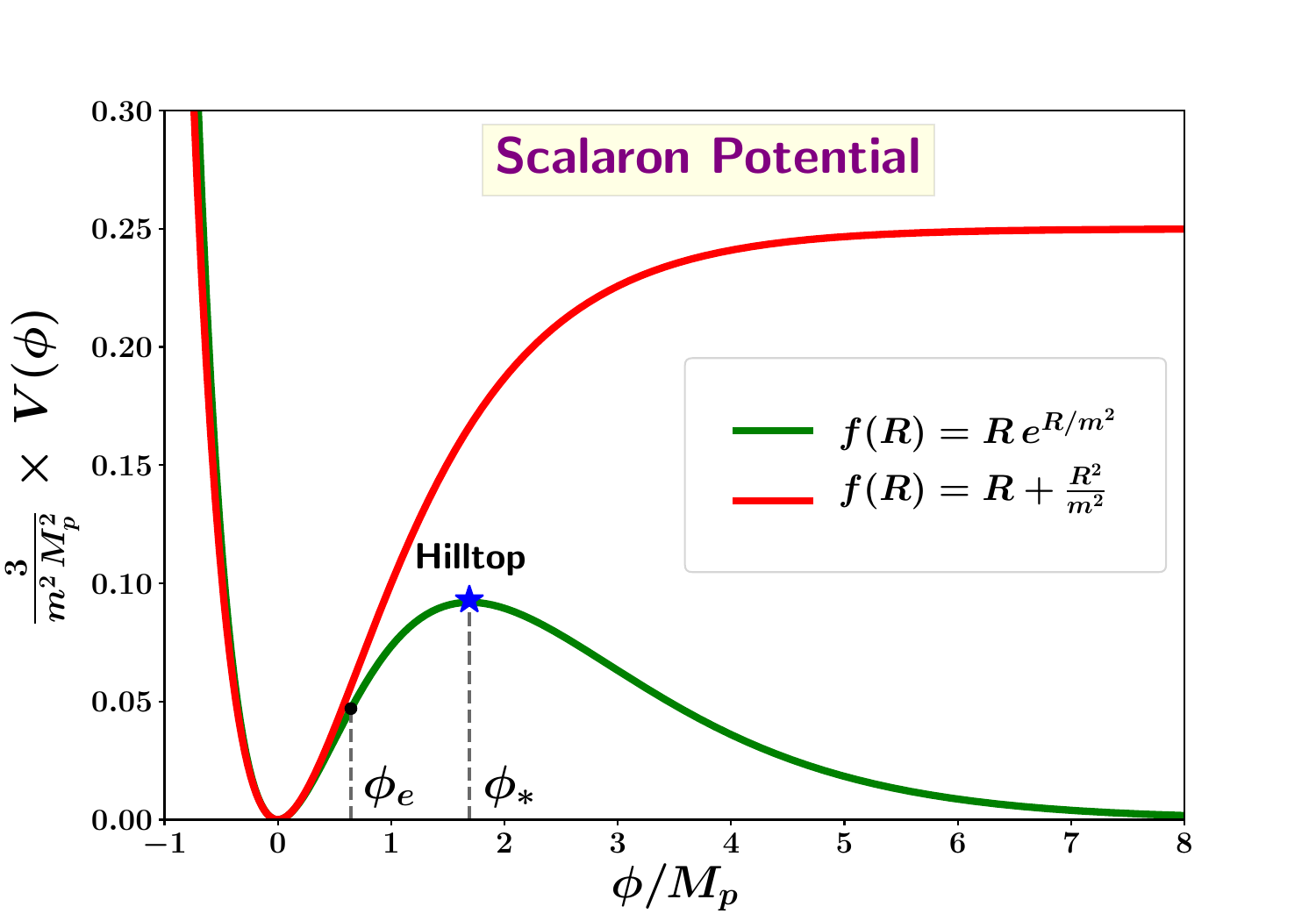}}
\subfigure[][]{
\includegraphics[width=0.6\textwidth]
{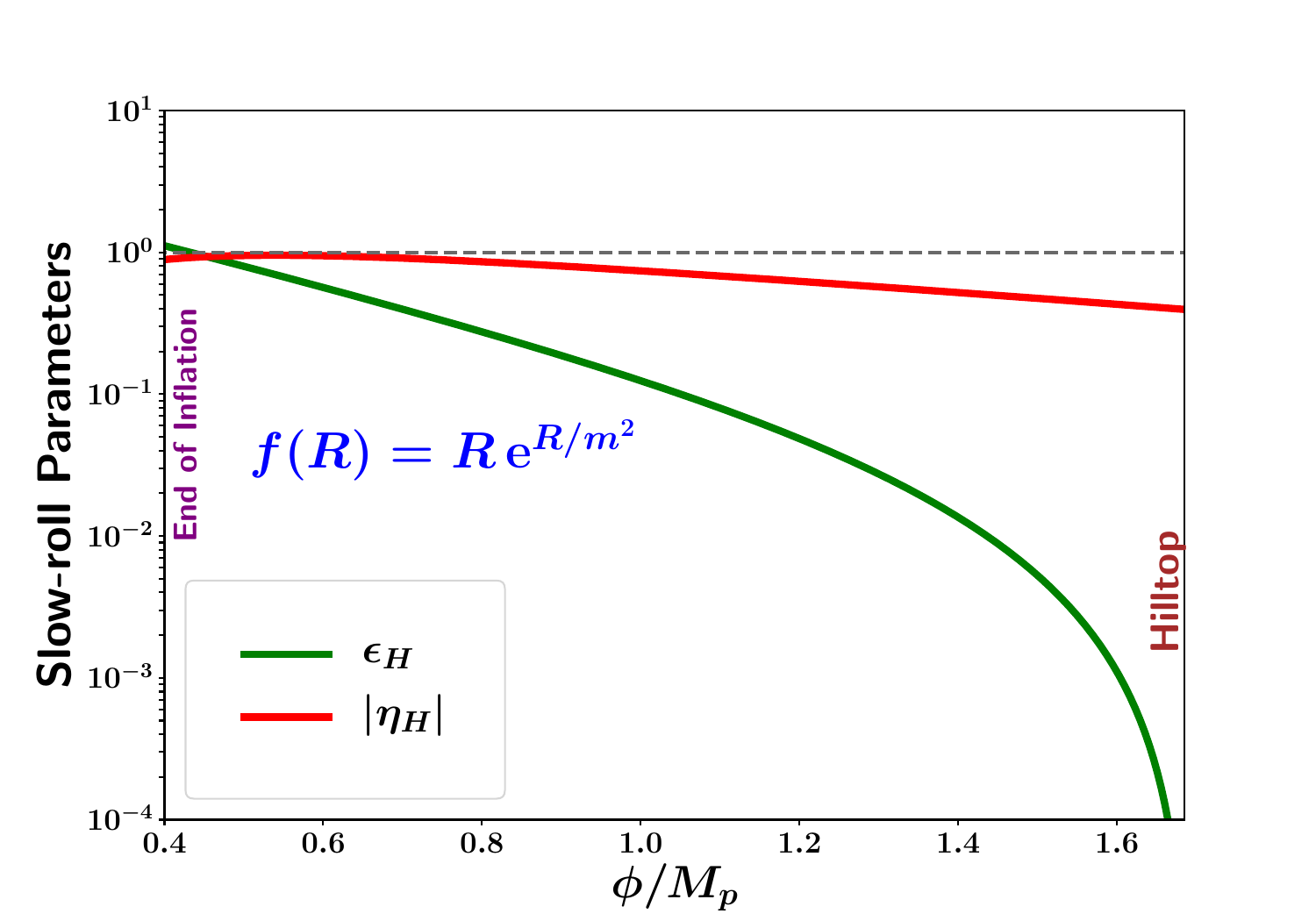}}
\caption{{\bf Top panel} shows the scalaron potential (\ref{eq:V_E_ReR}) for the model $f(R) =  R \, e^{R/m^2}$ in green colour (along with the scalaron potential (\ref{Vstar}) of the Starobinsky model $f(R) = R + R^2/m^2$ in red colour). The potential exhibits a maximum  for intermediate values of $\phi$ and a minimum at $\phi =0$, while it falls off exponentially for  $\phi \gg M_p$. The blue-colour star indicates the position of the CMB pivot scale $\phi_*$, which happens to be very close to the hilltop maximum, while $\phi_e$ marks the end of inflation. The slow-roll parameters $\epsilon_H$ and $|\eta_H|$ are  plotted in  the {\bf bottom panel}  in green and red colour curves respectively. This scalaron potential produces a highly red-tilted scalar power spectrum which does not satisfy the CMB constraints.} \label{fig:VE_ReR}
\end{figure}

The scalaron potential (\ref{eq:V_E_ReR}) is plotted in the top panel of Fig.~\ref{fig:VE_ReR}, along with the Starobinsky potential (\ref{Vstar}). The blue-colour star indicates the position of the CMB pivot scale $\phi_*$ which happens to be very close to the top of the hill. The  slow-roll parameters  $\epsilon_H$, $|\eta_H|$ are  plotted in the bottom  panel  in green and red colour curves respectively. While $\epsilon_H \ll 1$ at early times, its value is rapidly increasing leading to a relatively large value of the second slow-roll parameter $|\eta_H| \sim {\cal O}(1)$, as can be seen from the figure. Hence, slow-roll is not a very accurate approximation for this potential. Nevertheless, the slow-roll approximated value of $n_{_S}$ near the CMB pivot scale is $n_{_S} - 1 \simeq -0.8$ which indicates that the scalar power spectrum is highly red-tiled and hence this model is quite incompatible with the latest CMB 2$\sigma$ bound  $0.957 \leq \ns \leq 0.976$.

\subsection[Model \ $f(R) =  \frac{R}{1 - R/m^2}$]{Model \ $\boxed{ f(R) =  \frac{R}{1 - R/m^2}}$}
\label{sec:MoG_beta}

\begin{figure}[htb]
\centering
\subfigure[][]{
\includegraphics[width=0.6\textwidth]
{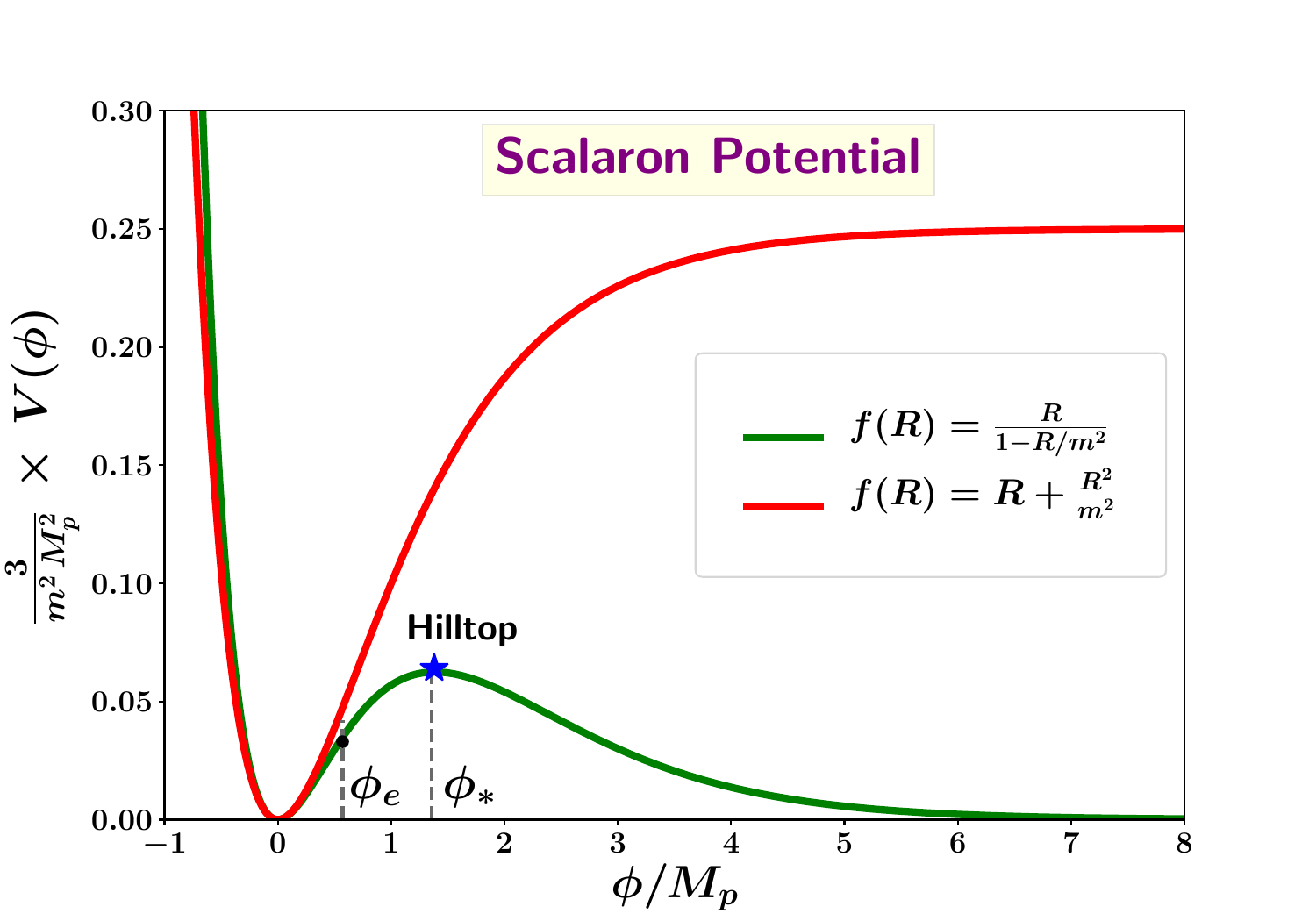}}
\subfigure[][]{
\includegraphics[width=0.6\textwidth]
{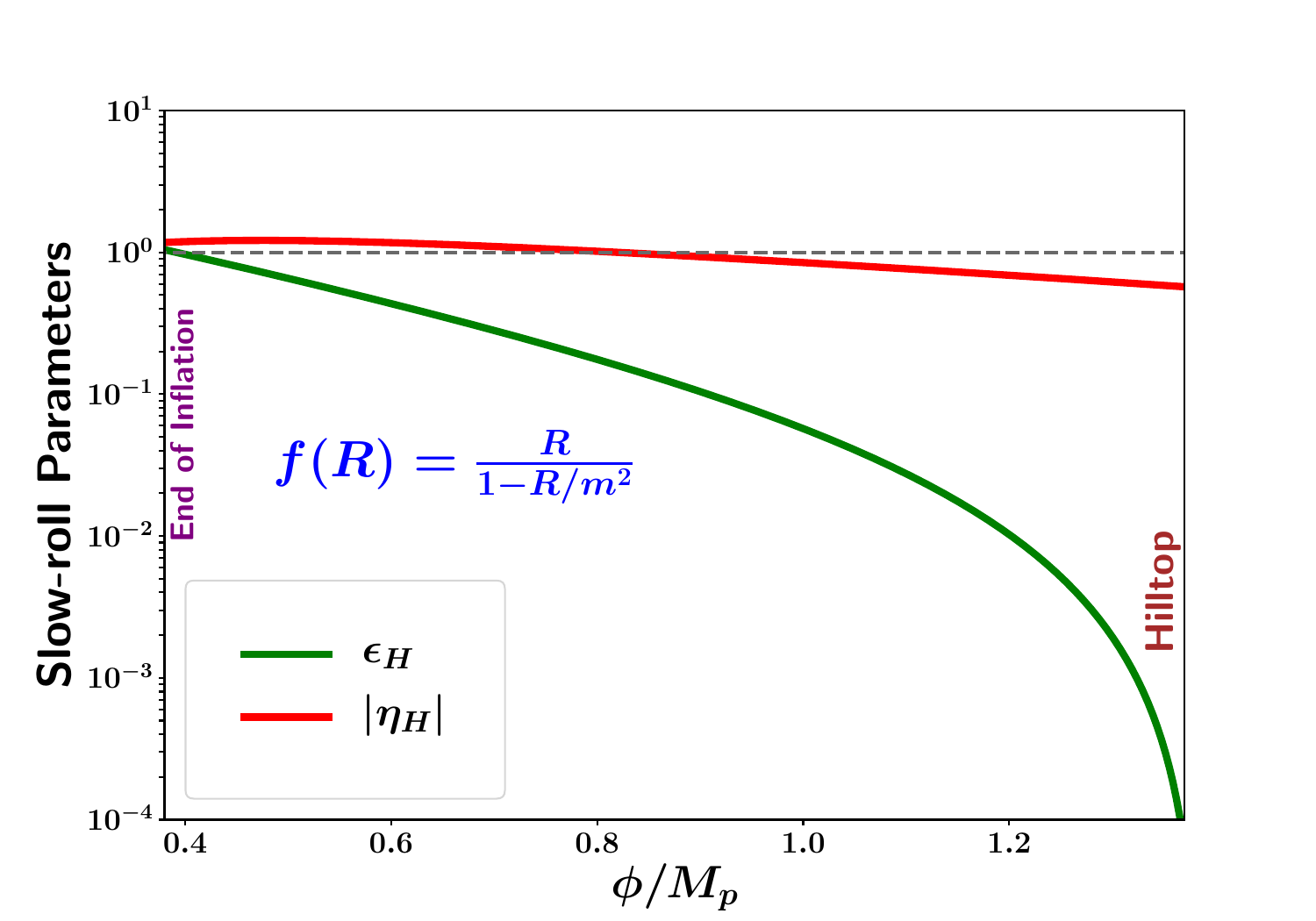}}
\caption{{\bf Top panel} shows the scalaron potential (\ref{eq:scalaron_pole}) for the model $f(R) =  \frac{R}{1 - R/R_m}$ in green colour (along with the scalaron potential (\ref{Vstar}) of the Starobinsky model $f(R) = R + R^2/m^2$ in red colour). The potential exhibits a maximum  for intermediate values of $\phi$ and a minimum at $\phi =0$, while it falls off exponentially for  $\phi \gg M_p$. The blue-colour star indicates the position of the CMB pivot scale $\phi_*$, which happens to be very close to the hilltop maximum, while $\phi_e$ marks the end of inflation. The slow-roll parameters $\epsilon_H$ and $|\eta_H|$ are  plotted in  the {\bf bottom panel}  in green and red colour curves respectively. This scalaron potential produces a highly red-tilted scalar power spectrum which does not satisfy the CMB constraints. } \label{fig:VE_MoG_beta}
\end{figure}

For model (\ref{power}), the scalaron potential for $\beta = 1$ is given by (\ref{eq:scalaron_pole}).  Setting $R_m = m^2$, we have
\begin{equation}\label{potrat}
V (\phi) =  \f{M_p^2 m^2}{3} e^{- \phi/M_p} \l( 1 - e^{-\phi/ 2 M_p} \r)^2 \, .
\end{equation}
This potential is illustrated in the top panel of Fig.~\ref{fig:VE_MoG_beta} along with the potential (\ref{Vstar}) of the Starobinsky model with the same value of $m$.  The blue-colour star indicates the position of the CMB pivot scale $\phi_*$ which happens to be very close to the top of the hill. 

The  slow-roll parameters  $\epsilon_H$, $|\eta_H|$ are  plotted in the bottom  panel  in green and red colour curves respectively. While $\epsilon_H \ll 1$ at early times, its value is rapidly increasing leading to a relatively large value of the second slow-roll parameter $|\eta_H| \sim {\cal O}(1)$, as can be seen from the figure (similarly to the case discussed for the potential  (\ref{eq:V_E_ReR})). Hence, slow-roll is not a very accurate approximation for this potential. Nevertheless, the slow-roll approximated value of $n_{_S}$ near the CMB pivot scale is $n_{_S} - 1 \simeq -1.3$ which indicates that the scalar power spectrum is highly red-tiled and hence this model is quite incompatible with the latest CMB 2$\sigma$ bound  $0.957 \leq \ns \leq 0.976$.

We observe that simple hilltop models with one mass scale $m$ fail to give
a reasonable value of the spectral index $n_S$. The reason for this is that the scalaron potential of such models has a maximum around which it behaves quadratically:
\begin{equation}
V (\phi) \approx V_m - \frac12 \mu^2 \left( \Delta \phi \right)^2 \, ,
\end{equation}
where $\Delta \phi = \phi - \phi_m$, and $\phi_m$ is the position of the maximum.  For potentials of the type under consideration, we have
\begin{equation}
V_m \sim 10^{-2} m^2 M_p^2 \, , \qquad \mu^2 \sim 10^{-1} m^2 \, .
\end{equation}
The pivot scale $\phi_*$ for such potentials happens to be very close to the maximum $\phi_m$, and the slow-roll parameters \eqref{eq:epsV_etaV} at this scale are then estimated as 
\begin{equation}
\epsilon_V \sim \frac{10 \left( \Delta \phi \right)^2}{M_p^2} \ll 1 \, , \qquad \eta_V \approx \left\{ \begin{array}{ll} 
- \dfrac{40}{9} \, , \quad \text{for potential \eqref{Vexpon} \, ,} \\[12pt]
- \dfrac{14}{3} \, , \quad \text{for potential \eqref{potrat} \, .}
\end{array}
\right.
\end{equation}
The parameter $| \eta_V |$ is thus of the order of a few, and the spectral index in the slow-roll approximation is estimated as $|n_S - 1| \approx 2 |\eta_V| \sim \text{a few}$, far off the value obtained from the CMB observations.

The next two models have two mass scales $m$ and $m_1$; they tend to the Starobinsky model as $m_1 \to \infty$ and, therefore, can fit observations very well.

\subsection[Model \ $f(R) =  R \, \frac{1+R/m^2}{1-R/m_1^2} $]{Model \ $\boxed{ f(R) =  R \, \frac{1+R/m^2}{1-R/m_1^2} } $}
\label{sec:star_ext}

For model (\ref{ration-scales}), the scalaron potential is given by (\ref{Vscales}): 

$$ V(\phi) = \frac{M_p^2 m_1^2}{3} \, e^{-2\phi/M_p} \, \left[ \, \sqrt{e^{\phi/M_p} + \frac{m_1^2}{m^2}} - \sqrt{1 + \frac{m_1^2}{m^2}}  \, \right]^2~.$$

\begin{figure}[ht]
\begin{center}
\includegraphics[width=0.85\textwidth]{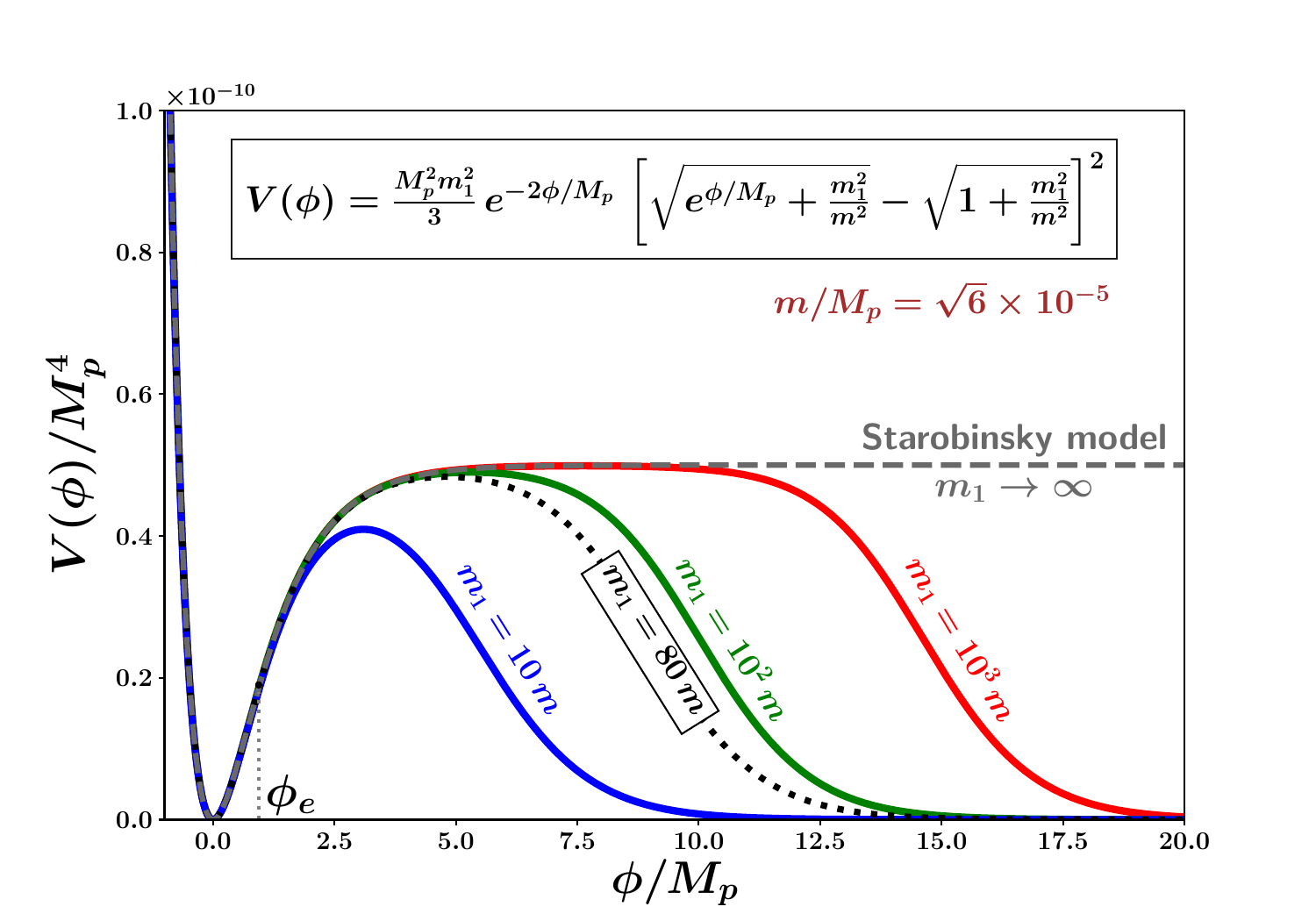}
\caption{This is a plot of the scalaron potential (\ref{Vscales}) for different values of the parameter $m_1$ with fixed $m$ (along with the scalaron potential of the Starobinsky model plotted in dashed grey colour). 
 Here, $\phi_e$ marks the end of inflation.
For $m_1 \gg m$, the potential begins to resemble a flattened 
hilltop at the CMB scale  $\phi = \phi_* \simeq 4.5 \, M_p$ and makes predictions for
CMB observables
which are identical to those of the Starobinsky-model for  $e^{\phi/M_p} \ll m_1^2/m^2$. 
Note that this model satisfies CMB constraints for $m_1 > 78\,m$.}
\label{fig:potE_hilltop_star_ext}
\end{center}
\end{figure}

For this potential, the CMB observables  $\lbrace n_{_S}, r \rbrace$ are sensitive to the ratio $m_1/m$ of the two mass scales in the Lagrangian (while the CMB normalisation fixes the value of $m$ or $m_1$ for a given ratio). For $m_1 \gg m$, the potential is almost identical to the potential of the Starobinsky model in the field range $e^{\phi/M_p} \ll m_1^2/m^2$. It falls off exponentially for  $e^{\phi/M_p} \gg m_1^2/m^2$. As we decrease the ratio $m_1/m$ from a large value towards unity, the potential starts to deviate from that of the Starobinsky model and slowly loses its 
tabletop feature. This is demonstrated in Fig.~\ref{fig:potE_hilltop_star_ext}.

\begin{figure}[htb]
\begin{center}
\includegraphics[width=0.85\textwidth]{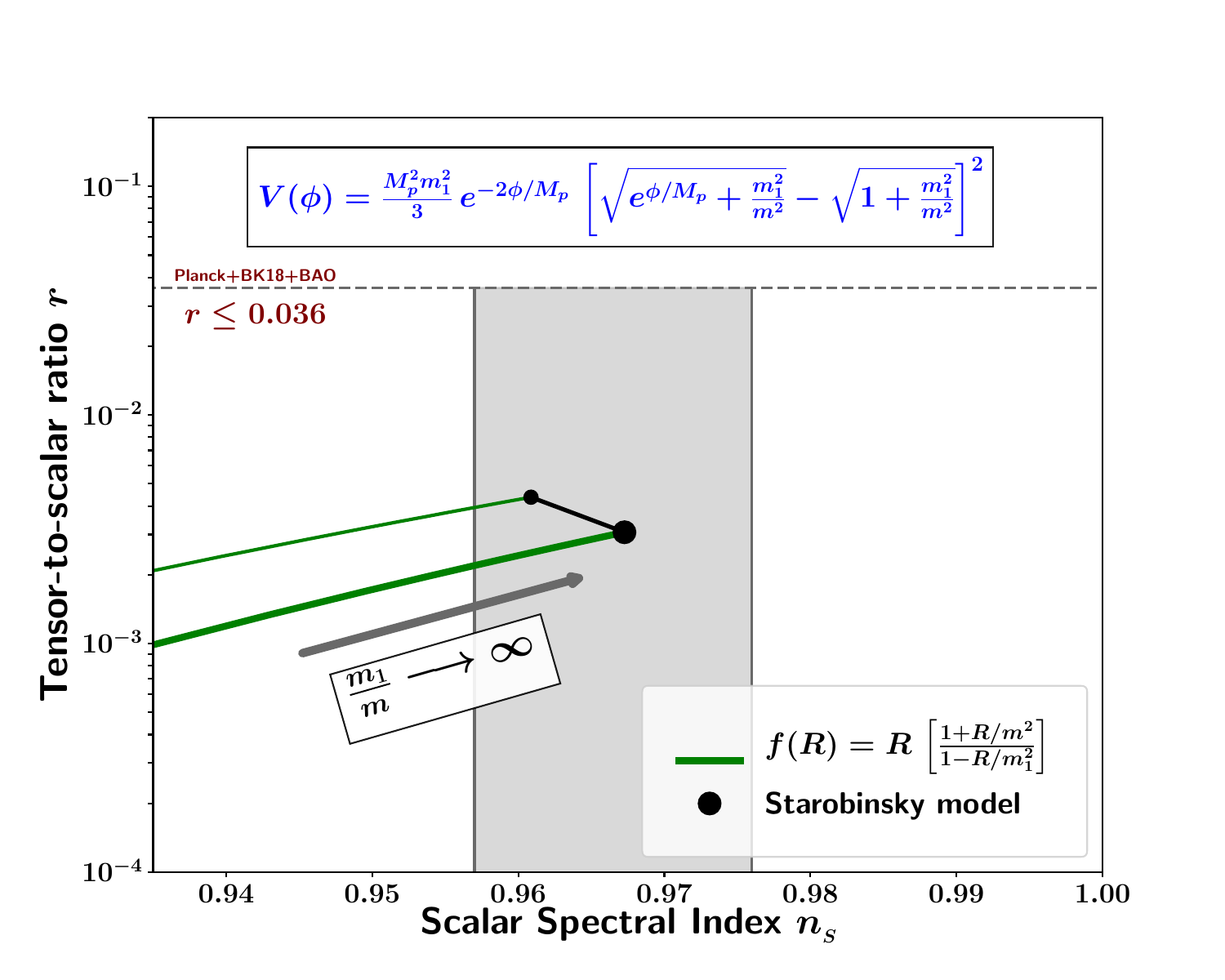}
\caption{This figure is a plot of  the  tensor-to-scalar ratio $r$ versus  the  scalar spectral index $\ns$    for  the model   (\ref{ration-scales}) with the scalaron potential (\ref{Vscales}) (the thinner and thicker   curves correspond to $N_* = 50,\, 60$ respectively).    The latest CMB 2$\sigma$ bound 
$0.957 \leq \ns \leq 0.976$ and the upper bound on the tensor-to-scalar ratio  $r\leq 0.036$ are indicated by the shaded grey colour region. Predictions of the model approach that of the Starobinsky model for $m_1 \geq  190 \, m$. Upon decreasing the ratio   $m_1/m$ in the  potential (\ref{Vscales}), the values of both $r$ and $n_{_S}$ decrease and eventually $n_{_S}$ becomes incompatible with the CMB data for $m_1 \leq 78 \,m$.}
\label{fig:ns_r_potE_star_ext}
\end{center}
\end{figure}

From Fig.~\ref{fig:potE_hilltop_star_ext}, we notice that when $m_1 < 80 \,m$ (roughly), the hilltop model starts to deviate substantially from the Starobinsky potential (in the region probed by the CMB, i.e.,  for  $\phi = \phi_* \simeq 4.5 \, M_p$). Our numerical analysis shows that  for $m_1 \geq 190 \,  m$, the   $\lbrace n_{_S},r \rbrace$ predictions of this model resemble those of the Starobinsky inflation, while  for $m_1 \leq 78 \,m$, the model becomes incompatible with the CMB data. The $\lbrace n_{_S},r \rbrace$ flow lines are illustrated in figure \ref{fig:ns_r_potE_star_ext}.

\subsection[Model  $ f(R) = R + \f{R^2}{m^2}+ \f{R^3}{m_1^4}$]{Model \ $\boxed{ f(R) = R + \f{R^2}{m^2} + \f{R^3}{m_1^4}  } $}
\label{sec:star_ext_cubic}

\begin{figure}[ht]
\begin{center}
\includegraphics[width=0.85\textwidth]{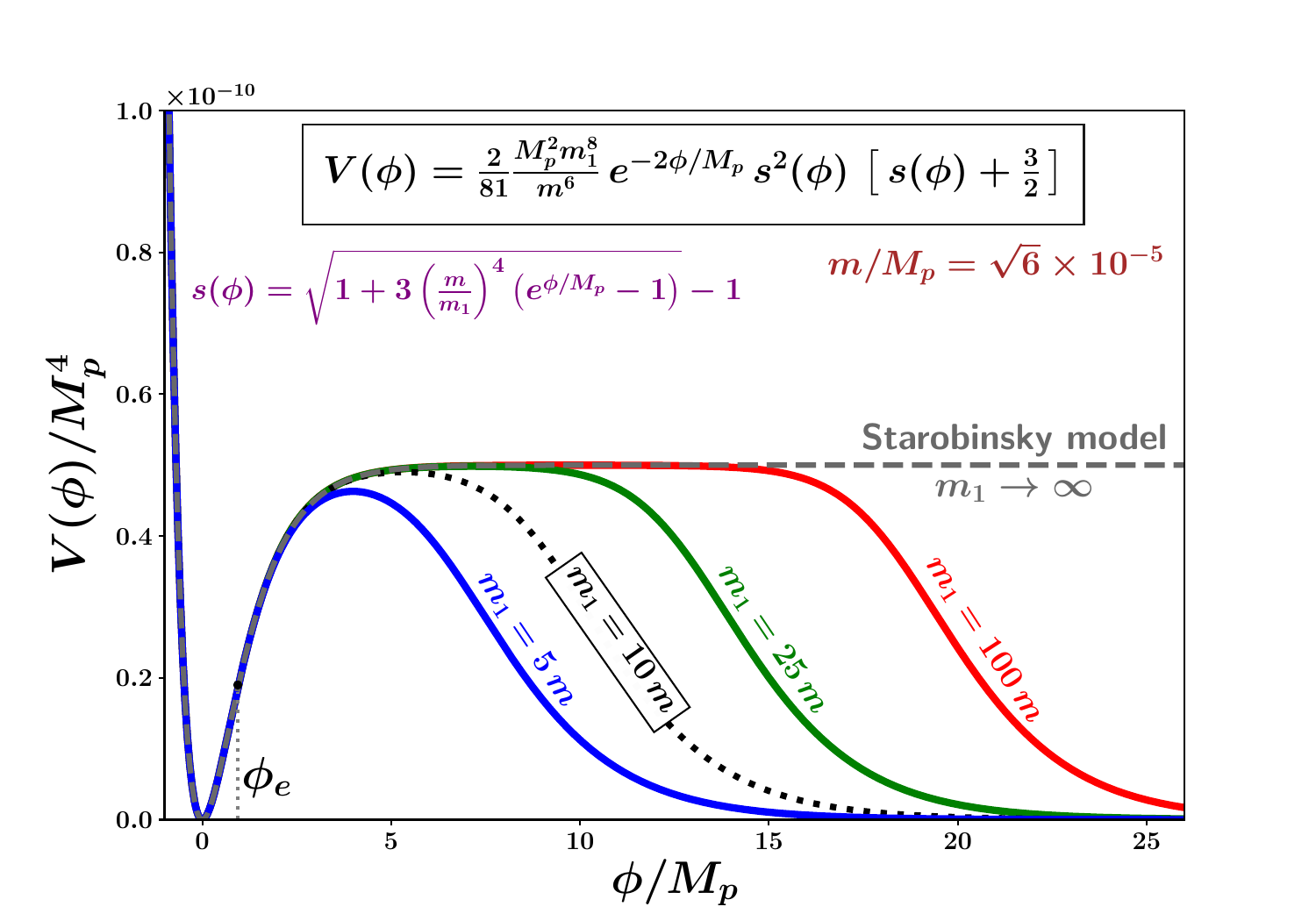}
\caption{This is a plot of the scalaron potential (\ref{cubicpot}) for different values of the parameter $m_1$ with fixed $m$ (along with the scalaron potential of the Starobinsky model plotted in dashed grey colour). 
 Here, $\phi_e$ marks the end of inflation.
For $m_1 \gg m$, the potential begins to resemble a flattened
hilltop at the CMB scale  $\phi = \phi_* \simeq 4.5 \, M_p$ and makes predictions for
CMB observables
which are identical to those of the Starobinsky-model for  $e^{\phi/M_p} \ll m_1^2/m^2$. Note that this model satisfies CMB constraints for $m_1 > 9\,m$.} 
\label{fig:potE_hilltop_star_ext_cubic}
\end{center}
\end{figure}

For model (\ref{cubic}), the scalaron potential is given by (\ref{cubicpot}): 
$$
V(\phi) = \frac{2}{81}\frac{M_p^2 m_1^8}{m^6} \, e^{-2\phi/M_p} \, s^2(\phi) \, \left[ \, s(\phi) + \frac{3}{2} \,  \right] \, ,
$$
with $s(\phi) = \sqrt{1 + 3m^4/m_1^4 \left( e^{\phi/M_p} - 1 \right) } - 1$. For this potential, the CMB observables  $\lbrace n_{_S}, r \rbrace$ are sensitive to the ratio $m_1/m$ of the two mass scales in the Lagrangian (while the CMB normalisation fixes the value of $m$ or $m_1$ for a given ratio). For $m_1 \gg m$, the potential exhibits an extended plateau and  
is almost  identical to the Starobinsky potential for small $\phi$ values.
 At large field values $V(\phi)$  falls off exponentially   
so that the resulting potential resembles a tabletop. As we decrease the ratio $m_1/m$ from a large value towards unity, the potential starts to deviate from the Starobinsky potential and 
looses its tabletop feature. This is demonstrated in Fig.~\ref{fig:potE_hilltop_star_ext_cubic}.

\begin{figure}[htb]
\begin{center}
\includegraphics[width=0.85\textwidth]{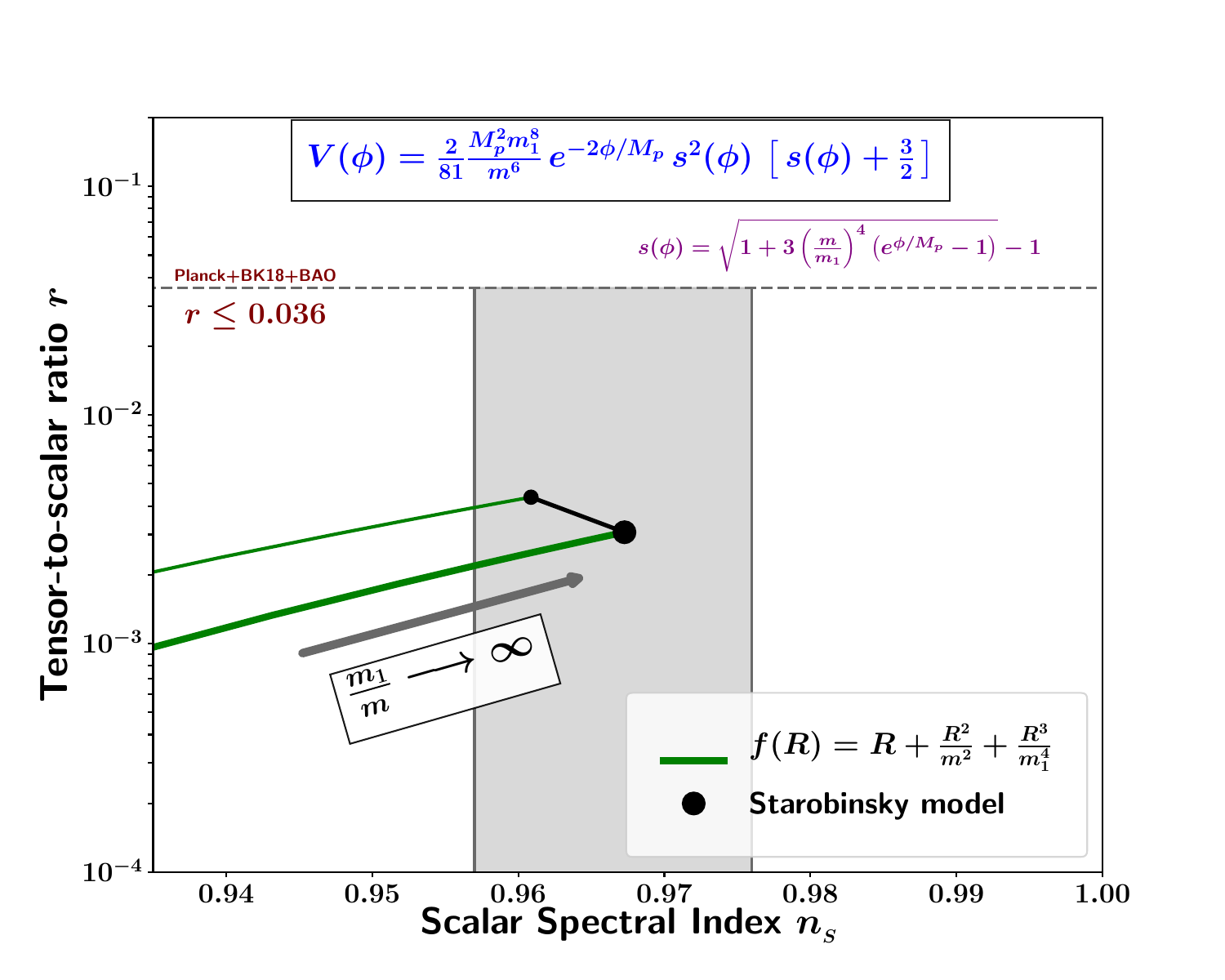}
\caption{This figure is a plot of  the  tensor-to-scalar ratio $r$ versus  the  scalar spectral index $\ns$    for  the model   (\ref{cubic}) with the scalaron potential (\ref{cubicpot}) (the thinner and thicker   curves correspond to $N_* = 50,\, 60$ respectively).    The latest CMB 2$\sigma$ bound 
$0.957 \leq \ns \leq 0.976$
 and
 the upper bound on the tensor-to-scalar ratio  $r\leq 0.036$ are
 indicated by the shaded grey colour region. Predictions of this model approach those of the Starobinsky model for $m_1 \geq 70\,  m$. Upon decreasing the ratio   $m_1/m$ in the  potential (\ref{cubicpot}), the values of both $r$ and $n_{_S}$ decrease and eventually  $n_{_S}$ becomes incompatible with the CMB data for $m_1 \leq 9 \,m$.}
\label{fig:ns_r_potE_star_ext_cubic}
\end{center}
\end{figure}

From Fig.~\ref{fig:potE_hilltop_star_ext_cubic}, we notice that when $m_1 < 10 \,m$ (roughly), the hilltop model starts to deviate substantially from the Starobinsky potential (in the region probed by the CMB, i.e.,  for  $\phi = \phi_* \simeq 4.5 \, M_p$). Our numerical analysis shows that, for $m_1 \geq 20 \, m$, the   $\lbrace n_{_S},r \rbrace$ predictions of this model resemble those of the Starobinsky inflation, while for $m_1 \leq 9 \,m$, the model becomes incompatible with the CMB data. The $\lbrace n_{_S},r \rbrace$ flow lines are illustrated in Fig.~\ref{fig:ns_r_potE_star_ext_cubic}. From Figs.\@ \ref{fig:ns_r_potE_star_ext} and \ref{fig:ns_r_potE_star_ext_cubic}, we can conclude that the  $\lbrace n_{_S},r \rbrace$ predictions of the MOG models  (\ref{cubicpot}) and  (\ref{Vscales}) are quite similar.  

We note that this model has also been considered previously in \cite{Berkin:1990nu, Faulkner:2006ub, Huang:2013hsb, Rodrigues-da-Silva:2021jab}, in which similar constraints on $m_1$ were obtained, namely, $m_1 > 6.8\, m$ \cite{Faulkner:2006ub}, $m_1 > 7.8\, m$ \cite{Huang:2013hsb}, and $m_1 > 9\, m$ \cite{Rodrigues-da-Silva:2021jab}. A small difference between the bound in \cite{Faulkner:2006ub} and our bound $m_1 > 9 \, m$ (coinciding with that of \cite{Rodrigues-da-Silva:2021jab}) is due to the shrinking of the allowed $n_{_S}$ values from WMAP to Planck.

From the above examples, we see that models with one mass scale typically fail to satisfy the observational CMB constraints while those with several mass scales, capable of 
incorporating a plateau/tabletop in the scalaron potential, can be compatible with observations just like the Starobinsky model. Regarding $f (R)$ as a general relativity theory with quantum corrections, the fact that higher powers of curvature should be sufficiently suppressed may be taken as an information about the nature of the underlying quantum field theory. This interesting issue lies beyond the scope of our paper.

\section{Evolution towards asymptotically vanishing gravity}
\label{sec:asymptote}

As we have already noted, the observable universe corresponds to the stable region 
around $\phi = 0$, which $\phi(t)$  reaches after rolling down the left slope of the 
hilltop/tabletop potential. However, the asymptotic region at $\phi \to \infty$ 
is also interesting since it corresponds to the region of asymptotically vanishing gravity.  
In this section, we will study the general features of this region.

\subsection{Asymptotic eternal inflation and Big-Rip}
\label{sec:inflation}

We first consider an important subclass of MOG theories, including (\ref{power}) and (\ref{ration-scales}), in which $f (R)$ monotonically blows up at $R = R_m = m^2$.  In this case, as we have seen earlier, the scalaron potential has the universal asymptotic form (\ref{scalexp}):
\begin{equation}\label{scalexp1}
V (\phi) \approx \frac{M_p^2 m^2}{3} e^{- \phi / M_p} \,, ~~~\phi \gg M_p~.
\end{equation}

The dynamical equations (\ref{eq:E_EOM}), (\ref{eq:E_Hubble}) in this case, neglecting the spatial curvature (which is justified in the asymptotic future) read
\begin{align}
&\ddot \phi + 3 \tilde H \dot \phi - \frac13 M_p m^2 e^{- \phi/M_p} = 0 \, , \label{eq:scalaron} \\
&\tilde H^2 = \frac{1}{2 M_p^2} \left( \frac12 \dot \phi^2 + \frac13 M_p^2 m^2 e^{- \phi / M_p} \right) \, , \label{eq:hubb} 
\end{align}
where $\tilde H \equiv \dot{\tilde a} / {\tilde a}$. It is convenient to proceed to the dimensional variables $\xi = \phi / M_p$ and $\tilde \tau = m \tilde t$.  
Equations (\ref{eq:scalaron}) and (\ref{eq:hubb}) then reduce to
\begin{align}
&\ddot \xi + 3 \tilde H \dot \xi - \frac13 e^{- \xi} = 0 \, , \label{scal-eq1} \\
&\tilde H^2 = \frac{1}{2} \left( \frac12 \dot \xi^2 + \frac13 e^{- \xi} \right) \, , \label{h-eq1}
\end{align}
where an overdot now denotes the derivative with respect to $\tilde \tau$. 

Next, we look for solutions of (\ref{scal-eq1}) and (\ref{h-eq1}) having the form 
\begin{equation}\label{dotxi}
\dot \xi = A e^{-\xi/2} \, ,
\end{equation}
which implies
\begin{equation}
e^{\xi/2} = \frac12 A \tilde \tau \, , \qquad \dot \xi = \frac{2}{\tilde \tau} \, ,
\label{eq:xi}
\end{equation}
where the integration constant has been absorbed by shifting the origin of time $\tilde \tau$.  Substituting (\ref{eq:xi}) into (\ref{scal-eq1}) and (\ref{h-eq1}), we obtain an algebraic equation for the constant $A$, with the solution $A = {1}/{2 \sqrt{3}}$. The Friedmann equation (\ref{h-eq1}) then reduces to
\begin{equation}\label{HE}
\tilde H = \frac{3}{\tilde \tau} \, , \qquad \tilde a (\tilde \tau) \propto \tilde \tau^3 \, .
\end{equation}

Note that, according to (\ref{dotxi}), we get the following equations in terms
 of the Einstein-frame time variable $\tilde t$ 
\begin{align}\label{slow-roll}
\dot \phi = \frac{M_p m}{2 \sqrt{3}} e^{- \phi / 2 M_p} \impl \dot \phi^2 = \frac14 V (\phi) \, , \quad \tilde H^2 = \frac{1}{2 M_p^2} \times \frac98 V (\phi) \, , \quad \dot \phi = - \frac{9}{8} \frac{ V ' (\phi)}{3 \tilde H} \, .
\end{align}
These are typical slow-roll inflation relations, with the replacement $V \to 9 V / 8$. Note that, in the asymptotic region $\phi \gg M_p$, according to (\ref{slow-roll}), 
\begin{equation}
\frac{\dot \phi }{\tilde H^2} \approx \frac{M_p^2}{\sqrt{V (\phi)}} \gg 1 \, ,
\end{equation}
so that the regular roll-down of the scalaron dominates over quantum diffusion. 

Proceeding to the Jordan frame is easy: the metric is simply multiplied by $e^{-\phi/M_p} = e^{- \xi}$.  The scale factor in the Jordan frame is, therefore,
\begin{equation}
a = e^{- \xi/2} \tilde a \propto \tilde \tau^2 \, .  
\end{equation}
Thus, the universe is expanding in the Jordan frame. The Jordan-frame time $\tau$ is related to the Einstein frame time $\tilde \tau$ via
\begin{equation}
d \tau  = e^{-\xi/2} d \tilde \tau = \frac{ 2 d \tilde \tau}{A \tilde \tau} \impl \tau = \frac{2}{A} \ln \tilde \tau \, .
\end{equation}
Hence, in the Jordan frame, one finds
\begin{equation}\label{aHJ}
a \propto \tilde \tau^2 \propto e^{A \tau} = e^{A m t} = e^{ m t / 2 \sqrt{3}}\, , \qquad H = \frac{m}{2 \sqrt{3}} \, .
\end{equation}
This is an asymptotically de~Sitter solution with the Hubble constant determined by $m$. {\em It describes eternal exponential inflation in the Jordan frame\/}.

It may be noted that our solution is future geodesically complete both in the Einstein and in the Jordan frame.

Next, let us consider general theories where $f (R)$ grows faster than any power as $R \to \infty$. An example of such a class of theories is given by \eqref{fexpow} with the asymptotics of the scalaron potential given by \eqref{Vexpow}:
\begin{equation} 
V (\phi) \approx \frac{M_p^2 m^2}{3} \left( \frac{\phi}{M_p} \right)^{2\omega} e^{- \phi / M_p} \, , \qquad \frac{\phi}{M_p} \gg 1 \, ,
\end{equation}
where we have denoted $2 \omega = 1/\beta$. Analogues of \eqref{scal-eq1} and \eqref{h-eq1} now read
\begin{align}
&\ddot \xi + 3 \tilde H \dot \xi - \frac13 \xi^{2\omega} e^{- \xi} = 0 \, , \label{scal-eq2} \\
&\tilde H^2 = \frac{1}{2} \left( \frac12 \dot \xi^2 + \frac13 \xi^{2\omega} e^{- \xi} \right) \, . \label{h-eq2}
\end{align}
We can look for asymptotic solutions of (\ref{scal-eq2}) and (\ref{h-eq2}) having the form 
\begin{equation}\label{dotxi2}
\dot \xi = A \xi^\omega e^{-\xi/2} \, ,
\end{equation}
which asymptotically implies
\begin{equation}
\xi^{- \omega} e^{\xi/2} = \frac12 A \tilde \tau \, , \qquad \dot \xi = \frac{2}{\tilde \tau} \, .
\end{equation}
From \eqref{scal-eq2}, the algebraic equation for the constant $A$ remains to be the same as before, with the result $A = 1/2 \sqrt{3}$. Solution \eqref{HE} in the Einstein frame also remains intact:
\begin{equation}
\tilde H = \frac{3}{\tilde \tau} \, , \qquad \tilde a (\tilde \tau) \propto \tilde \tau^3 \, .
\end{equation}
However, the scale factor in the Jordan frame now reads
\begin{equation}
a = e^{- \xi/2} \tilde a \propto \tilde \tau^2 \xi^{- \omega} \approx \tilde \tau^2 \left( \ln \tilde \tau^2 \right)^{- \omega} \, .  
\end{equation}
The Jordan-frame time $\tau$ is related to the Einstein frame time $\tilde \tau$ via
\begin{equation}\label{texpow}
d \tau  = e^{-\xi/2} d \tilde \tau = \frac{ 2 d \tilde \tau}{A \tilde \tau} \xi^{- \omega} = \frac{ d \ln \tilde \tau^2}{A} \left( \ln \tilde \tau^2 \right)^{- \omega} \, .
\end{equation}

For $\omega < 1$, in the Jordan frame, one finds
\begin{align}
&\tau = \frac{\left( \ln \tilde \tau^2 \right)^{1 - \omega}}{A (1 - \omega)} \, , \qquad \tilde \tau^2 = e^{\left[ A (1 - \omega) \tau \right]^{1/(1 - \omega)}} \, , \\[3pt]
&a \propto \tilde \tau^2 \left( \ln \tilde \tau^2 \right)^{- \omega} \propto e^{\left[ A (1 - \omega) \tau \right]^{1/(1 - \omega)}} \tau^{- \omega / ( 1 - \omega)} \, . \label{agen}
\end{align}
For $\omega = 0$, we obtain the previous result \eqref{aHJ}. For theory \eqref{fexpon} with the asymptotics of the scalaron potential given by \eqref{exponent}, we have $\omega = 1/2$, and our asymptotical solution \eqref{agen} describes a super-exponential inflation: $a \propto e^{(A \tau / 2)^2}/\tau$.

For $\omega = 1$, in the Jordan frame, from \eqref{texpow} we find
\begin{align}
&\tau = \frac{1}{A} \ln \ln \tilde \tau^2 \, , \qquad \tilde \tau^2 = e^{e^{A \tau}} \, , \\[3pt]
&a \propto \tilde \tau^2 \left( \ln \tilde \tau^2 \right)^{- 1} \propto e^{e^{A \tau} - A \tau} \, . 
\end{align}
This is a double-exponential inflation.

For $\omega > 1$ (which corresponds to $\beta < 1/2$ in \eqref{fexpow}), the integral of $\tau$ from \eqref{texpow} converges, and we obtain a Big-Rip in the Jordan frame, with the scale factor blowing up in a finite time $\tau$. 

\subsection{Potentials with a general asymptotic exponent}
\label{sec:asym}

The previous subsection focussed on a potential having the exponential asymptotic form $V (\phi) \propto e^{- \phi / M_p}$. We now examine the case when (for large positive or negative values of $\phi$) the scalaron potential has the form 
\begin{equation}\label{asb}
V (\phi) \approx \frac12 M_p^2 m^2 e^{- \gamma \phi / M_p} \, , \qquad \gamma > 0 \, , \quad \gamma \ne 1 \, .
\end{equation} 
Instead of (\ref{eq:scalaron}) and (\ref{eq:hubb}), we now have the following Einstein-frame equations:
\begin{align}
&\ddot \phi + 3 \tilde H \dot \phi - \frac{\gamma}{2} M_p m^2 e^{- \gamma \phi/M_p} = 0 \, , \label{scal-b} \\
&\tilde H^2 = \frac{1}{4 M_p^2} \left( \dot \phi^2 + M_p^2 m^2 e^{- \gamma \phi / M_p} \right) \, . \label{h-b}
\end{align}
Introducing $\xi = \phi / M_p$ and $\tilde \tau = m \tilde t$, we obtain analogues of equations (\ref{scal-eq1}) and (\ref{h-eq1}):
\begin{align}
&\ddot \xi + 3 \tilde H \dot \xi - \frac{\gamma}{2} e^{- \gamma \xi} = 0 \, , \label{scal-b1} \\
&\tilde H^2 = \frac{1}{4} \left( \dot \xi^2 + e^{- \gamma \xi} \right) \, . \label{h-b1}
\end{align}

Again, we look for solutions of the form 
\begin{equation} \label{solxib}
\dot \xi = A e^{- \gamma \xi / 2}  \impl e^{\gamma \xi / 2} = \frac{\gamma}{2} A \tilde \tau \, , \quad \dot \xi = \frac{2}{\gamma \tilde \tau} \, .
\end{equation}
Then
\begin{equation}\label{solhb}
\tilde H^2 = \frac{1}{\gamma^2 \tilde \tau^2} \left( 1 + \frac{1}{A^2} \right)  \, .
\end{equation}
Substituting (\ref{solxib}) and (\ref{solhb}) into (\ref{scal-b1}), we obtain an algebraic equation for $A$: 
\begin{equation}
\left( 1 + \frac{1}{A^2} \right)  = \frac{9}{\gamma^2} \, .
\end{equation}
Solution in this form exists only for $\gamma < 3$. In this case, one finds the Einstein-frame expansion rate to be
\begin{equation}\label{solga}
\tilde H = \frac{3}{\gamma^2 \tilde \tau} \impl \tilde a ( \tilde \tau) \propto \tilde \tau^{3 / \gamma^2} \, .
\end{equation}
For $\gamma < \sqrt{3}$,  equation (\ref{solga}) describes power-law inflation while, for $\gamma > \sqrt{3}$, one gets decelerated expansion. Note that, in both cases, the universe is future geodesically complete in the Einstein frame.\footnote{Asymptotic solution \eqref{solga} was obtained in \cite{Rodrigues-da-Silva:2021jab} for the cubic model \eqref{cubic}, in which case $\gamma = 1/2$ [see \eqref{ascub}].} 

In the Jordan frame, the metric is multiplied by $e^{- \phi/M_p} = e^{- \xi}$.  The scale factor in the Jordan frame is, therefore,
\begin{equation}
a = e^{- \xi/2} \tilde a \propto \tilde \tau^{(3 - \gamma)/\gamma^2}\, .  
\end{equation}
The Jordan-frame time $\tau$ is related to the Einstein frame time $\tilde \tau$ via
\begin{equation}
d \tau = e^{-\xi/2} d \tilde \tau \propto \tilde \tau^{- 1/\gamma} d \tilde \tau \impl \tau \propto \left\{ \begin{array}{cl} \tilde \tau^{(\gamma - 1)/\gamma} \, , & \gamma > 1 \, , \smallskip \\ \tilde \tau_0^{-(1 - \gamma)/\gamma} - \tilde \tau_{}^{-(1 - \gamma)/\gamma} \, , & \gamma < 1 \, . \end{array} \right.
\end{equation}
Hence,
\begin{equation}
a \propto \tilde \tau^{(3 - \gamma)/\gamma^2} \propto \left\{ \begin{array}{cl} \tau^{(3 - \gamma)/\gamma (\gamma - 1)} \, , & \gamma > 1 \, , \smallskip \\ \left( \tau_0 - \tau \right)^{- (3 - \gamma)/\gamma (1 - \gamma)} \, , & \gamma < 1 \, . \end{array} \right.
\label{scale_Jordan}
\end{equation}

Thus, for $1 < \gamma < 3$, the universe is expanding eternally in the Jordan frame, while, for $\gamma < 1$, it runs into a `Big-Rip' singularity in a finite Jordan time.
The Big-Rip occurs at $\tau = \tau_0$ when the expansion factor $a(\tau)$ diverges,
as does the Hubble parameter (and its derivatives). 

$\bullet$ As an example, consider evolution on the left-hand slope $\phi < 0$ of the potential (\ref{Vstar}) of the Starobinsky model, where it has an approximate form
\begin{equation}
V (\phi) \approx \frac12 M_p^2 m^2 e^{- 2 \phi / M_p} \, .
\end{equation} 
This corresponds to $\gamma = 2$ of the general case considered above.   Thus, expansion on this slope proceeds as $\tilde a ( \tilde t) \propto \tilde t^{3/4}$ in the Einstein frame, which is not inflation since $\ddot {\tilde a} < 0$. In the Jordan frame, 
one finds $a \propto t^{1/2}$.

$\bullet$ Another example is provided by potential (\ref{pwpot}) for integer $n$. The asymptote of the potential at $\phi/M_p \gg 1$ is given by (\ref{asb}) with $\gamma =1 - 1/n$ for $n > 1$. Since $\gamma < 1$, the universe encounters a future `Big-Rip' singularity in the Jordan frame. This can easily be seen from (\ref{scale_Jordan}), which reduces to ($\tau \equiv t$)
\begin{equation}
a \propto \left( t_0 - t \right)^{- (3 - \gamma)/\gamma (1 - \gamma)} = \left( t_0 - t \right)^{- n (2 n + 1)/(n - 1) } \, ,
\label{PLI}
\end{equation}
and demonstrates that $a, H, {\dot H}$, $R \to \infty$ as $t\to t_0$. Equation (\ref{PLI}) is in agreement with the results of \cite[Eq.~(28)]{Carloni:2004kp} and \cite[Eq.~(26)]{Ivanov:2011np}.

Since $\gamma = 1$ in the MOG models (\ref{Vstar}), (\ref{exponent}), (\ref{eq:scalaron_pole}) and (\ref{Vscales}), these models will exhibit eternal inflation in the asymptotic region $\phi \to \infty$. In models \eqref{alpha}, (\ref{pwpot}) and (\ref{cubicpot}) on the other hand, $\gamma < 1$, implying that the universe will run into a `Big-Rip' singularity at $\phi \to \infty$. Note that behaviour \eqref{asb} with $\gamma > 1$ is not encountered in stable theories in the asymptotic region of $\phi \to \infty$ as is clear from the asymptotic expressions \eqref{ualpha} and \eqref{alpha}.

\section{Summary}
\label{sec:summary}

$f (R)$ gravity contains one extra degree of freedom (the scalaron $\phi$), which was used to generate inflation in the Starobinsky model \cite{Starobinsky:1980te, Vilenkin:1985md}. This model contains only linear (Einstein) and quadratic terms in the scalar curvature, $f (R) = R + R^2 / m^2$, and the corresponding scalaron potential in the Einstein frame has one extremum (stable minimum) at $\phi = 0$, with a plateau extending to infinity in $\phi$-space. We have shown that $f (R)$ models containing only one free parameter, usually translate into hilltop potentials in the Einstein frame, which are ruled out by current CMB observations. The Starobinsky potential appears to be the sole exception to this rule. 

The presence of additional terms of higher order in $R$ typically results in the scalaron potential acquiring a flattened hilltop or tabletop form, provided the additional terms enter with relatively small coefficients. An example is given by $f (R) = R + R^2 / m^2 + R^3 / m_1^4$ with $m_1 \gg m$. In such a case, evolution towards the stable minimum at $\phi = 0$ (corresponding to $R = 0$ in the Jordan frame) proceeds in a manner similar to that in the Starobinsky model, yielding primordial perturbations which are in perfect agreement with CMB observations. In the other limiting case when $m_1 \simeq m$ this potential turns into a hilltop and comes into conflict with CMB constraints. Our paper provides several new examples of $f(R)$ inflation whose potential in the Einstein frame resembles a tabletop and which satisfy CMB observations. (The stability condition $f'' (R) > 0$ is complied with in all of our models.) 

If $f (R)$ gravity is regarded as a general relativity theory with quantum corrections producing inflation, then the fact that higher powers of curvature in this model should be sufficiently suppressed may be taken as an information about the nature of the underlying quantum field theory. This is relevant also for non-local extensions of $f(R)$ gravity \cite{Koshelev:2016xqb, Koshelev:2022olc}.

If the universe is quantum created with the scalaron at the top of the hilltop/plateau, 
then the subsequent evolution of $\phi(t)$ can proceed either towards the stable region 
with $\phi = 0$ (to the left of the hill) or towards
 the asymptotic region $\phi \to \infty$ (to the right of the hill).  
In case of the latter, the subsequent evolution of the universe will be sensitive to
the behaviour of the scalaron potential in the asymptotic region $\phi \to \infty$, 
which, in turn, depends on the behaviour of $f (R)$ at large values of $R$.
In this case the 
universe can expand in two distinct ways:
(i) either
 towards a Big-Rip singularity,
at which $H,{\dot H} \to \infty$, or (ii) towards eternal inflation. 
Eternal exponential inflation is characteristic of all models in which $f (R)$ diverges at a finite value of the scalar curvature $R = R_m$.  Big-Rip singularity is characteristic of models which grow as power law $f (R) \propto R^{1+\alpha}$ with $\alpha > 1$ or exponentially with a slow exponent \eqref{fexpow} with $\beta < 1/2$ as $R \to \infty$.  In all such models, gravitational interaction of matter and gravitons vanishes $(G_{\rm eff} \to 0$) in the Jordan frame in the corresponding limit. 

Finally we would like to draw attention to the fact that if a closed universe is quantum created in the vicinity of a field value that corresponds to the Hubble-exit of  CMB scales, then the effects  of a positive spatial curvature  will  potentially be imprinted  on the low multipole angular power spectrum of CMB \cite{CMB_closed1,CMB_closed2}. In this case, quantum fluctuations during inflation must be analysed in a spatially closed background in order to determine the properties of primordial power spectra. While such initial conditions may be somewhat fine-tuned, they will nevertheless have important cosmological implications. Moreover, the curvature term could also become relevant at late times (closer to the present epoch) depending on the details of reheating  and the post-inflationary dynamics of the universe. Both the primordial and the late time effects of the spatial curvature 
could have important consequences for the low multipole CMB power spectra. We plan to revert  to this  issue in a future work.

\section*{Acknowledgements}
The authors are grateful to Oliver Gould for raising the question about the graviton interaction in the Jordan frame. SM and VS admire the perseverance and dedication shown by Yuri Shtanov who worked on this paper in Kiev at the time of a grave national crisis precipitated by the war in Ukraine. Y.\,S. is supported by the National Academy of Sciences of Ukraine under project 0121U109612 and by the Taras Shevchenko National University of Kyiv under project 22BF023-01. V.\,S. was partially supported by the J.\,C.~Bose Fellowship of Department of Science and Technology, Government of India. S.\,S.\,M. is supported as an STFC Consolidated Grant [Grant No. ST/T000732/1].

\appendix

\section{Two-particle scattering in the Jordan frame}
\label{sec:scatter}

Consider the process of scattering of two particles on the background of the metric $g_{\mu\nu}$ in the Jordan frame, or on the background of the corresponding metric $\tilde g_{\mu\nu}$ and scalaron $\phi$ in the Einstein frame. As usual, we assume the wavelengths of the scattering particles to be much smaller than the curvature of the background metric or the variation scale of the scalaron field. The metrics in two frames are related by \eqref{conform}, \eqref{phi}:
\begin{equation} \label{met}
\tilde g_{\mu\nu} = e^{\phi / M_p} g_{\mu\nu} \, , \qquad \tilde g^{\mu\nu} = e^{- \phi / M_p} g^{\mu\nu} \, .
\end{equation}

Two-particle scattering is characterised by cross-section, which is the area of an effective two-surface which both particles should cross in order to scatter. Such an effective two-surface has the same {\em coordinate\/} extension in two frames, but its area is measured by the corresponding metric in each frame.  From this observation, in view of the first equation in \eqref{met}, we immediately get
\begin{equation}\label{sig}
\tilde \sigma ( \tilde \cE ) = e^{\phi / M_p} \sigma (\cE) \, ,
\end{equation} 
where $\sigma$ and $\tilde \sigma$ are the cross-sections, and $\cE$ and $\tilde \cE$ are the collision energies of the same process measured in the Jordan and Einstein frames, respectively.

To relate the collision energies in two frames, we note that a particle is represented by a wave packet, in which its kinematics is encoded in a phase factor $e^{\ri S (x)}$ with a rapidly varying phase $S (x)$. This phase will be the same in two frames (it reflects the coordinate propagation of the particle wave packet, which is frame-independent). Hence, the covariant components of the particle four-momentum $p_\mu = \nabla_\mu S (x)$ are the same in two frames. 

Let the four-momenta of the scattering particles be $p_{1 \mu}$ and $p_{2 \mu}$. We then have, using the second relation in \eqref{met},
\begin{equation}\label{E}
\tilde \cE^2 = 2 p_{1 \mu} p_{2 \nu} \tilde g^{\mu\nu} = e^{- \phi / M_p} 2 p_{1 \mu} p_{2 \nu} g^{\mu\nu} =  e^{- \phi / M_p} \cE^2 \, .
\end{equation}
From \eqref{sig} and \eqref{E}, we finally obtain
\begin{equation}\label{crossrel}
\sigma (\cE) = e^{- \phi / M_p} \tilde \sigma \left( e^{- \phi / 2 M_p} \cE \right) \, .
\end{equation}
This equation is valid for scattering of particles of any nature, and allows one to relate the corresponding cross-sections in the Einstein and Jordan frames.

Consider now scattering of two gravitons. The corresponding cross-section in the Einstein frame is determined by the Einstein gravity theory, and its behaviour for collision energies well below the Planckian energy is known:
\begin{equation}\label{gr}
\tilde \sigma (\cE) \sim G^2 \cE^2 \, .
\end{equation}
Hence, in the Jordan frame, we will have
\begin{equation}\label{result}
\sigma (\cE) = e^{- 2 \phi / M_p} \tilde \sigma (\cE) \sim e^{- 2 \phi / M_p} G^2 \cE^2 = G_\text{eff}^2\, \cE^2 \, .
\end{equation}
We observe that the graviton scattering cross-section in the Jordan frame is determined by the effective gravitational coupling $G_\text{eff} = e^{- \phi / M_p} G = G / f' (R)$, vanishing in the asymptotic region of $\phi \to \infty$, corresponding to $f' (R) \to \infty$.

One can arrive at the same conclusion by considering the expansion of the $f (R)$ Lagrangian in the Jordan frame around the background solution with $R = R_0$:
\begin{equation}
f (R) = f (R_0 + \delta R) = f(R_0) + f' (R_0) \delta R + \ldots \, .
\end{equation}
Here, $\delta R$ contains graviton perturbations to all orders. Confining ourselves to the term linear in $\delta R$, we see that interaction of gravitons will proceed as in general relativity theory with the effective coupling $G_\text{eff} = G / f' (R_0)$.  In view of \eqref{gr}, we again obtain our result \eqref{result}.

Similar cross-sections \eqref{gr}, \eqref{result} will be obtained for the graviton-scalaron scattering.  As regards scattering of two scalarons with potential \eqref{asb} (with unrestricted $\gamma$) in the asymptotic region, its cross-section due to quartic self-interaction in the Einstein frame will be of the order
\begin{equation}
\tilde \sigma (\cE) \sim \frac{m^4}{M_p^4 \cE^2} e^{- 2 \gamma \phi / M_p} \, .
\end{equation}
In the Jordan frame, we, therefore, will have from \eqref{crossrel}:
\begin{equation}
\sigma (\cE) = \tilde \sigma (\cE) \sim \frac{m^4}{M_p^4 \cE^2} e^{- 2 \gamma \phi / M_p} \, ,
\end{equation}
from where we find that the cross-section in both frames vanishes in the asymptotic region when $\phi \to \infty$ and $G_{\rm eff} \to 0$.

\section{Scalaron potential for models with odd $f (R)$}
\label{sec:odd}

Some MOG
models could have non-unique solutions of equation (\ref{dir1}), with values of the associated
field $\Omega$ being
restricted for such solutions. In this case an equivalent potential in the Einstein frame can be obtained by `gluing' the Einstein frame solutions
on several intervals of the Jordan scalar curvature $R$.

\begin{figure}[ht]
\begin{center}
\includegraphics[width=.495\textwidth]{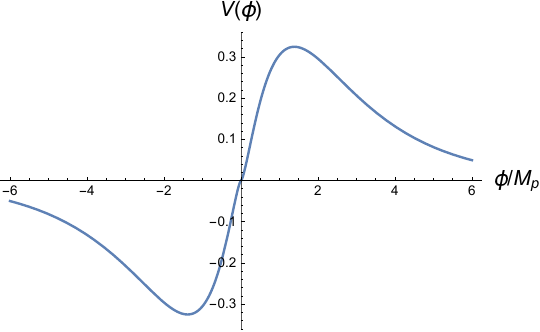} \hfill \includegraphics[width=.495\textwidth]{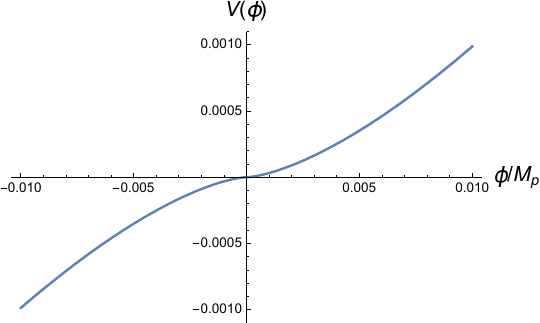}
\caption{Potential (\ref{R3pot}) in units $2 M_p^2 m^2 / 9$. The right plot zooms the central region so that the extremal point at $\phi = 0$ becomes visible. \label{fig:R^3}}
\end{center}
\end{figure}

This will indeed be the case for all $f (R)$ Lagrangians which are odd in $R$. The function $f' (R)$ is then even, and equation (\ref{dir1}) has at least two solutions.  The simplest example is
\begin{equation}
f (R) = R + \frac{R^3}{3 m^4} \, .
\end{equation}
Equation (\ref{dir1}) and its solutions in this case read
\begin{equation}
1 + \frac{R^2}{m^4} = \Omega \, , \qquad R = \pm m^2 \sqrt{ \Omega - 1} \, , 
\end{equation}
where the signs $\pm$ refer to the respective regions $R \gtrless 0$. The Legendre transform (\ref{dir2}) and potential (\ref{W}) in these regions are given by
\begin{equation}
h (\Omega) = \pm \frac{2 m^2}{3 \Omega^2} (\Omega - 1)^{3/2} \, , \qquad W (\Omega) = \pm \frac{2 M_p^2 m^2}{9 \Omega^2} (\Omega - 1)^{3/2} \, .
\end{equation}
In both field regions, the value of $\Omega$ is restricted to $\Omega \geq 1$, so that the range of the scalaron $\phi = M_p \ln \Omega$ is $\phi \geq 0$. These two regions can be combined by reversing the sign of the scalaron field in the region $R < 0$ and by gluing it with the scalaron in the region $R > 0$.  We obtain the potential for the scalaron with unrestricted values in the form
\begin{equation}\label{R3pot}
V (\phi) = \frac{2 M_p^2 m^2}{9} \text{sign} \left( \phi \right)\, e^{- 2 |\phi|/M_p} \left( e^{|\phi|/M_p} - 1 \right)^{3/2} \, .
\end{equation}
This potential is continuously differentiable everywhere. It's plot in units $2 M_p^2 m^2 / 9$ is presented in Fig.~\ref{fig:R^3}.

Similar potentials will be obtained for other odd functions $f (R)$ such as
\begin{equation}\label{harmon}
f (R) = R \sec \frac{R}{m^2} \, , \qquad f (R) = m^2 \tan \frac{R}{m^2} \, .
\end{equation}
It is interesting that the critical point $R = 0$ is not stable for such potentials. The stable point corresponds to a negative value of the scalar curvature and hence to anti-de~Sitter space.  Moreover, the two theories \eqref{harmon} will have an absolute limiting curvature $| R | < R_m = \pi m^2 / 2$.


\end{document}